%% file: main.tex
\documentclass[10pt, conference]{IEEEtran}

\input{includes/includes}

\input{includes/pygments}
\input{includes/macros}

\input{includes/lst3d}
\input{includes/lstfstar}
\input{includes/lstsmt2}
\usepackage{tabularray}
\usepackage[turnon]{notes}
\usepackage{listings}

\usepackage{graphicx} 

\let\ls\lstinline
\newcommand\lstt[1]{\lstinline[language=3D]$#1$}
\newcommand\lsf[1]{\lstinline[language=fstar]$#1$}
\newcommand\lss[1]{\lstinline[language=SMT2]$#1$}

\newcommand\ourtool{\textsc{3dGen}\xspace}

\def\BibTeX{{\rm B\kern-.05em{\sc i\kern-.025em b}\kern-.08em
    T\kern-.1667em\lower.7ex\hbox{E}\kern-.125emX}}
\begin{document}

\title{3DGen: AI-Assisted Generation of\\ Provably Correct Binary Format Parsers}

\author{\IEEEauthorblockN{Sarah Fakhoury,
Markus Kuppe, Shuvendu K. Lahiri, Tahina Ramananandro and
Nikhil Swamy}
\IEEEauthorblockA{Microsoft Research, Redmond, USA\\
\{sfakhoury, makuppe, shuvendu, taramana, nswamy\}@microsoft.com}}
\maketitle
\nolinenumbers
\input{body/0.abstract}


\input{body/1.intro}

\input{body/2.background}
\input{body/3.3DGen}

\input{body/4.Dataset}

\input{body/5.results}

\input{body/related}

\input{body/7.conclusion}

\bibliographystyle{IEEEtran}
\bibliography{main}

\newpage
\appendix
\input{body/8.Appendix}

\end{document}

%% file: includes/includes.tex
\usepackage{tabularx}
 \usepackage{booktabs}
\usepackage[font=small,labelfont=bf,tableposition=top]{caption}
\usepackage{hyperref}
\usepackage{boldline}
\usepackage{xcolor}
\usepackage{tabularx}
\usepackage{tcolorbox}
\usepackage{color, colortbl}
\usepackage{bigstrut}
\usepackage{amsmath,amsfonts}
\usepackage{amsmath}
\usepackage{array}
\usepackage{algorithmic}
\usepackage{balance}
\usepackage{blindtext}
\usepackage{caption}
\usepackage{hyperref}
\usepackage[noabbrev]{cleveref}
\usepackage{cite}
\usepackage{comment}
\usepackage{enumitem}
\usepackage{eqparbox}
\usepackage{fancybox}
\usepackage{fancyvrb}
\usepackage{framed}
\usepackage{graphicx}
\usepackage{ifthen}
\usepackage{listings} 
\usepackage{mathrsfs}
\usepackage{mdwmath}
\usepackage{mdwtab}
\usepackage{multirow}
\usepackage{pifont}
\usepackage{stfloats}
\usepackage{textcomp}
\usepackage{times}
\usepackage{url}
\usepackage{xspace}
\usepackage[normalem]{ulem}
\usepackage[framemethod=TikZ]{mdframed}
\usepackage{caption}
\usepackage{subcaption}
\usepackage{tabularx}
\usepackage[switch]{lineno}
\usepackage{algorithm}

\usepackage{titlesec}

\usepackage{tcolorbox}
\tcbuselibrary{listings,skins}
 
\usepackage{mathtools}
\usepackage{blindtext}
\usepackage{lstautogobble}

\DeclareGraphicsExtensions{.pdf,.jpeg,.png}
\graphicspath{{images/}}

\titlespacing\section{0pt}{8pt plus 4pt minus 2pt}{4pt plus 2pt minus 2pt}

\newcommand{\rom}[1]{\uppercase\expandafter{\romannumeral #1\relax}}

\definecolor{gray50}{gray}{.5}
\definecolor{gray40}{gray}{.6}
\definecolor{gray30}{gray}{.7}
\definecolor{gray20}{gray}{.8}
\definecolor{gray10}{gray}{.9}
\definecolor{gray05}{gray}{.95}
\definecolor{gray01}{gray}{.97}

\newlength\Linewidth
\def\findlength{\setlength\Linewidth\linewidth
\addtolength\Linewidth{-4\fboxrule}
\addtolength\Linewidth{-3\fboxsep}
}
\newenvironment{examplebox}{\par\begingroup
  \setlength{\fboxsep}{3pt}\findlength
  \setbox0=\vbox\bgroup\noindent
  \hsize=0.90\linewidth
  \begin{minipage}{0.90\linewidth}\normalsize}
    {\end{minipage}\egroup
    \textcolor{gray20}{\fboxsep1.2pt\fbox
     {\fboxsep5pt\colorbox{gray05}{\normalcolor\box0}}}
    \endgroup\par\noindent
    \normalcolor\ignorespacesafterend}

\usetikzlibrary{shadows}
\usepackage{graphics}
\newmdenv[
    tikzsetting= {fill=blueish},
    skipabove=0.33em,
    skipbelow=0.33em,
    linewidth=1pt,
    innerleftmargin=4pt,
    innerrightmargin=4pt,
    innertopmargin=2pt,
    innerbottommargin=2pt,
    linecolor=gray95,
    roundcorner=2pt, 
    shadow=true,
    shadowsize=4pt,
    shadowcolor=gray95
]{questionbox}

\newmdenv[
    tikzsetting= {fill=greenish},
    skipabove=0.33em,
    skipbelow=0.33em,
    linewidth=1pt,
    innerleftmargin=4pt,
    innerrightmargin=4pt,
    innertopmargin=2pt,
    innerbottommargin=2pt,
    linecolor=gray95,
    roundcorner=2pt, 
    shadow=true,
    shadowsize=4pt,
    shadowcolor=gray95
]{answerbox}

\newmdenv[
    skipabove=0.33em,
    skipbelow=0.33em,
    innerleftmargin=4pt,
    innerrightmargin=4pt,
    innertopmargin=2pt,
    innerbottommargin=2pt,
]{lessonbox}

\usepackage{tikz}

\newenvironment{lesson}
{
    \begin{lessonbox}
    Lessons Learned:\\
}
{\end{lessonbox}}

\newenvironment{result}
{\begin{answerbox}}
{\end{answerbox}}

\newenvironment{question}
{\begin{questionbox}}
{\end{questionbox}}

\definecolor{javared}{rgb}{0.6,0,0} 
\definecolor{javagreen}{rgb}{0.25,0.5,0.35} 
\definecolor{javapurple}{rgb}{0.5,0,0.35} 
\definecolor{javadocblue}{rgb}{0.25,0.35,0.75} 

\lstdefinestyle{basejava}{
  language=java,
  showstringspaces=false,
  basicstyle=\scriptsize\ttfamily,
  keywordstyle=\bfseries\color{javapurple},
  commentstyle=\itshape\blue,
  identifierstyle=\blue,
  frame=none,
  backgroundcolor=\color{white},
}

\lstdefinestyle{CustomJava}{
  belowcaptionskip=\baselineskip,
  breaklines=true,
  xleftmargin=\parindent,
  language=java,
  showstringspaces=false,
  basicstyle=\scriptsize\ttfamily,
  keywordstyle=\bfseries\color{javapurple},
  commentstyle=\itshape\blue,
  identifierstyle=\blue,
  belowskip=1pt,
  frame=shadowbox,
  backgroundcolor=\color{gray01},
  gobble=0
}

\lstdefinestyle{codit}{
  belowcaptionskip=\baselineskip,
  breaklines=true,
  language=java,
  showstringspaces=false,
  basicstyle=\scriptsize\ttfamily,
  keywordstyle=\bfseries\color{javapurple},
  commentstyle=\itshape\blue,
  identifierstyle=\blue,
}

\newtcblisting{mylisting}[2][]{
    arc=3mm,
    listing only, 
    listing style=codit,
    title=#2,
    #1
    }

%% file: includes/pygments.tex
\newcommand\blue[1]{\textcolor[rgb]{0.00,0.00,1.00}{{#1}}}

\definecolor{blueish}{RGB}{250, 250, 255}
\definecolor{greenish}{RGB}{250, 255, 250}
\definecolor{redish}{RGB}{255, 200, 200}

\definecolor{gray05}{gray}{0.95}
\definecolor{gray08}{gray}{0.92}
\definecolor{gray10}{gray}{0.90}
\definecolor{gray12}{gray}{0.88}
\definecolor{gray15}{gray}{0.85}
\definecolor{gray18}{gray}{0.82}
\definecolor{gray20}{gray}{0.80}
\definecolor{gray25}{gray}{0.75}
\definecolor{gray30}{gray}{0.70}
\definecolor{gray35}{gray}{0.65}
\definecolor{gray40}{gray}{0.60}
\definecolor{gray45}{gray}{0.55}
\definecolor{gray50}{gray}{0.50}
\definecolor{gray55}{gray}{0.45}
\definecolor{gray60}{gray}{0.40}
\definecolor{gray65}{gray}{0.35}
\definecolor{gray70}{gray}{0.30}
\definecolor{gray75}{gray}{0.25}
\definecolor{gray80}{gray}{0.20}
\definecolor{gray85}{gray}{0.15}
\definecolor{gray90}{gray}{0.10}
\definecolor{gray95}{gray}{0.05}

\definecolor{addition}{rgb}{0,0.1,0.5}
\definecolor{dkblue}{rgb}{0,0.1,0.5}
\definecolor{dkgreen}{rgb}{0,0.4,0}
\definecolor{dkred}{rgb}{0.6,0,0}
\definecolor{dkpurple}{rgb}{0.7,0,1.0}
\definecolor{purple}{rgb}{0.9,0,1.0}
\definecolor{olive}{rgb}{0.4, 0.4, 0.0}
\definecolor{teal}{rgb}{0.0,0.4,0.4}
\definecolor{azure}{rgb}{0.0, 0.5, 1.0}
\definecolor{gray}{rgb}{0.5, 0.5, 0.5}
\definecolor{dkgrey}{rgb}{0.2, 0.2, 0.2}
\definecolor{lilac}{rgb}{0.70, 0.04, 0.08}
\definecolor{applegreen}{rgb}{0.55, 0.71, 0.0}

%% file: includes/macros.tex
\usepackage{amssymb}

\usepackage{tikz}
\newcommand{\tikzxmark}{%
\tikz[scale=0.23] {
    \draw[line width=0.7,line cap=round] (0,0) to [bend left=6] (1,1);
    \draw[line width=0.7,line cap=round] (0.2,0.95) to [bend right=3] (0.8,0.05);
}}

\xspace%

\newcommand\tshark{Wireshark\xspace}
\newcommand{\threedsymdiff}{\textsc{3dTestGen}\xspace}

\newcommand\fstar{F$^\star$\xspace}

\newcommand\maybecolor[1]{\color{#1}}
\newcommand{\Comment}[1]{}

\newcommand{\citep}[1]{\cite{#1}}

\newtcbox{\inlinebox}[1][]{enhanced,
 box align=base,
 nobeforeafter,
 colback=blueish,
 size=small,
 left=0pt,
 right=0pt,
 boxsep=2pt,
 #1}

\newcommand{\RS}[2]{%
    {\small
    \begin{result}
        \textbf{\hyperref[rq-#1]{Result {#1}}:~}{\emph {#2}}%
    \end{result}}
}

\renewcommand{\cref}[1]{\Cref{#1}}

\newcommand{\rqa}{Can 3DGen produce format specifications for networks protocols standardized in RFCs?}

\newcommand{\rqb}{How do candidate format specifications generated for the same protocol differ?}

\newcommand{\rqc}{How do format specifications generated by \ourtool compare to handwritten specifications?}

\newcommand{\threedGen}{\textsc{3DGen}\xspace}

\newcommand{\SC}{\textsc{3DSynChk}}

\newcommand{\Ex}{\textsc{3DExec}}
\newcommand{\DslDoc}{\it 3DDoc}

\newcommand{\ADSLProgGen}{\threedGen}
\newcommand{\candProgs}{\textit{CandProgs}}
\newcommand{\queryLLM}{{\textsc{QueryLLM}}}

\newcommand{\symbTestGen}{\threedsymdiff}

\newcommand{\isSat}{\textsc{SATModel}}

\newcommand{\seedTests}{\it I_0}
\newcommand{\posSeedTests}{\seedTests^+}
\newcommand{\negSeedTests}{\seedTests^-}
\newcommand{\posTests}{I^+}
\newcommand{\negTests}{I^-}
\newcommand{\labelTest}{\textsc{LabelInputs}}
\newcommand{\Req}{\textit{RFC}}

%% file: body/0.abstract.tex
\begin{abstract}
Improper parsing of attacker-controlled input is a leading source of software security vulnerabilities, especially when programmers transcribe informal format descriptions in RFCs into efficient parsing logic in low-level, memory unsafe languages. Several researchers have proposed formal specification languages for data formats from which efficient code can be extracted. However, distilling informal requirements into formal specifications is challenging and, despite their benefits, new, formal languages are hard for people to learn and use.

In this work, we present 3DGen, a framework that makes use of AI agents to transform mixed informal input, including natural language documents (i.e.,  RFCs) and example inputs into format specifications in a language called 3D. To support humans in understanding and trusting the generated specifications, 3DGen uses symbolic methods to also synthesize test inputs that can be validated against an external oracle. Symbolic test generation also helps in distinguishing multiple plausible solutions. Through a process of repeated refinement, 3DGen produces a 3D specification that conforms to a test suite, and which yields safe, efficient, provably correct, parsing code in C.

We have evaluated 3DGen on 20 Internet standard formats, demonstrating the potential for AI-agents to produce formally verified C code at a non-trivial scale. A key enabler is the use of a domain-specific language to limit AI outputs to a class for which automated, symbolic analysis is tractable. 

\end{abstract}

%% file: body/1.intro.tex
\section{Introduction}
\label{sec:intro}

Improper parsing of attacker-controlled input is a leading source of software security vulnerabilities,\footnote{\url{https://cwe.mitre.org/data/definitions/20.html}}\footnote{\url{https://cwe.mitre.org/data/definitions/502.html}} especially when programmers transcribe informal format descriptions into efficient parsing logic in low-level, memory unsafe languages. For example, the format of TCP headers is specified in natural language and packet diagrams in the classic RFCs 793 and 9293; meanwhile, \textit{tcp\_input.c}, the TCP header parser in the Linux kernel was patched to prevent an out of bounds access in 2019, after being in the kernel for nearly 20 years. 

In response, researchers have proposed languages for describing low-level binary message formats backed by code generators that yield parsing and serialization tools, e.g., Nail~\cite{nail14osdi} and EverParse\cite{everparse19usenix,everparse3d22pldi}. EverParse is notable in that it produces formally verified C code from a format description language (called 3D), guaranteeing memory safety, functional correctness, and double-fetch freedom..  

In an ideal world, one might hope for specifications to always be written in domain-specific languages (DSLs) like 3D that yield trustworthy executable code. However, more commonly, specifications are not entirely formal and come from a variety of sources, ranging from natural language documents, diagrams, example code snippets, sample input/output pairs, etc. Extracting a formal specification from such a variety of sources requires a significant human effort, typically requiring a process that involves:

\begin{enumerate}
\item Learning a new DSL;
\item Understanding the informal specification;
\item Expressing one's understanding of the informal specification in the DSL;
\item Iterating to refine intent, revisiting the previous steps to arrive at a desired specification.
\end{enumerate}

This is challenging enough that developers often directly transcribe informal specifications into executable code in general purpose programming languages, leaving the door open to low-level coding errors that lead to security vulnerabilities. 

\subsection{\ourtool: A Framework for AI-assisted DSL Programming}

\begin{figure}
    \centering
    \includegraphics[width=0.9\columnwidth]{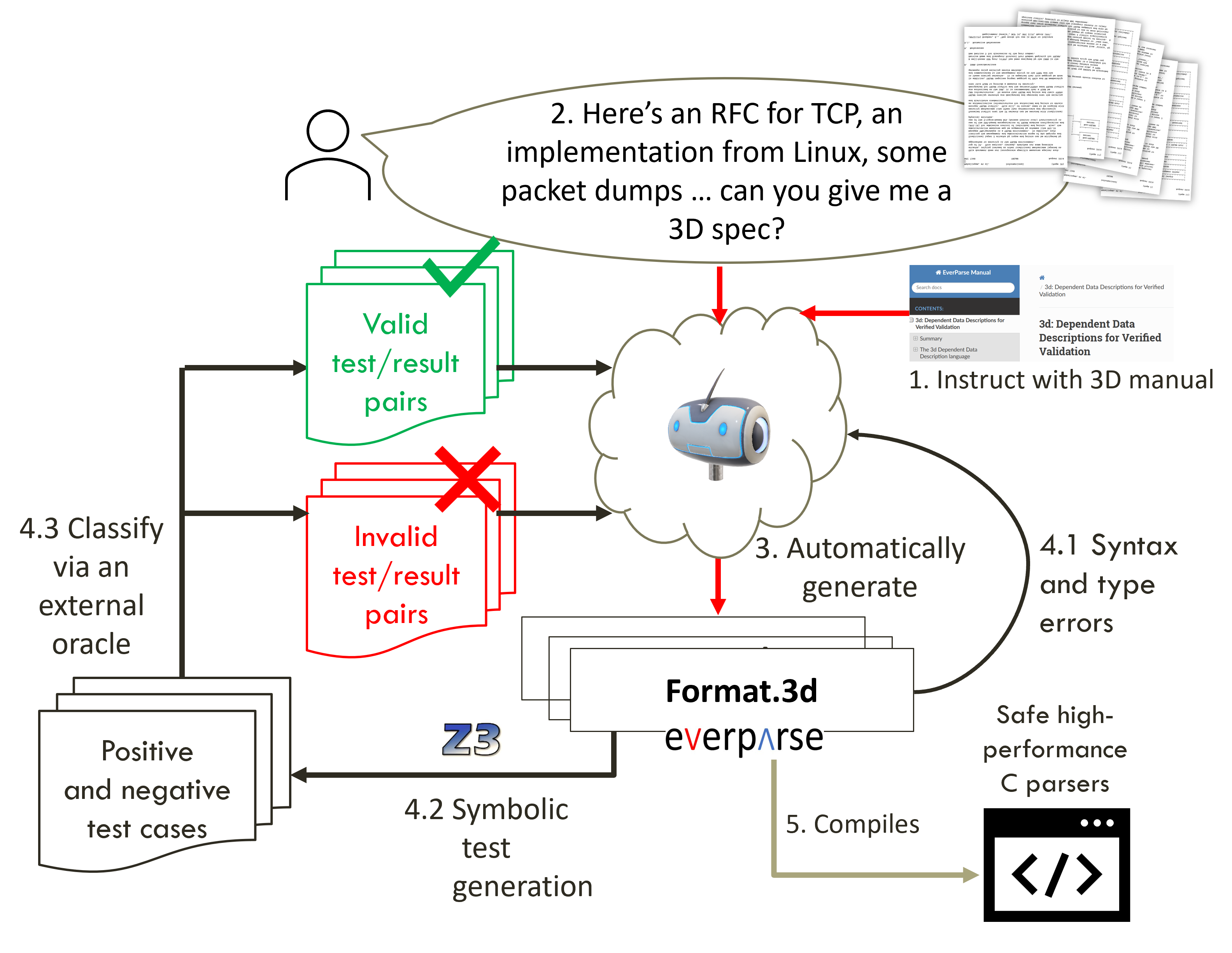}
    \caption{Workflow of \ourtool}
    \label{fig:mainworkflow}
\end{figure}

In this work, we present \ourtool, a framework that uses
AI agents to assist a human in translating an informal specification to executable code via a DSL, grounded specifically
in generating binary format parsers using 3D.
Our framework is agnostic to the AI model used, though for our experiments we use GPT-4~\cite{openai2023gpt4}. The core of \ourtool is an automated intent-refinement loop which assists a user in constructing a 3D specification that matches an oracle's behavior on a set of test inputs. Figure~\ref{fig:mainworkflow} sketches the high-level workflow, whose main elements mirror the steps outlined above. 

\smallskip\noindent\emph{1. Teaching a DSL to an agent:} 3D is a small language whose syntax is based on C's syntax for typedefs, structures, and unions. Its manual is relatively compact, consisting of around 2,000 lines of text and around 20 examples. We ``teach" \ourtool about 3D by giving it access to the manual, and the ability to query parts of the manual based on techniques that we describe in \S\ref{subsec:3dagent}. 

\smallskip\noindent\emph{2 \& 3. Digesting informal specifications into 3D:} A user gathers a collection of informal specifications, including  natural language documents that describe message formats and sample test inputs, and presents it to \ourtool. In turn, our framework, prompts the underlying agents to generate a 3D specification. 

\smallskip\noindent\emph{4. Refining intent} The 3D compiler analyzes candidate 3D specifications, providing syntax and type errors that we feed back to the agents to repair their code until the produced 3D specifications are at least well defined. Next, exploiting the fact that 3D is a small language with a well-designed formal semantics, we develop \threedsymdiff a new symbolic test-case generator for 3D. This allows us to automatically generate new test inputs for candidate specifications, and we rely on various external oracles to decide the intended classification of an input. Enriching the input with the new test cases, we repeat the loop.

Having converged with \ourtool's help on a given 3D specification, the user relies on EverParse to generate verified C code that is guaranteed to parse only all messages that are well-formed according to the specification.

We evaluate \ourtool on 20 Internet standard formats, starting from their specification in RFCs. While \threedsymdiff can generate tests from candidate specifications, we still require oracles to label those tests with their desired outcome; specifically, we use the \tshark\footnote{https://www.wireshark.org/docs/man-pages/tshark.html} network packet analyzer to 
decide if a packet should be accepted or not. Interestingly, in 11 cases, \ourtool discovers constraints specified in RFCs that \tshark does not enforce. Additionally, for 7 protocol formats, we also evaluated \ourtool's ability to produce 3D specification that match the behavior of prior, handwritten specifications for various formats provided as samples by the authors of EverParse. In this setting, \ourtool uncovers 3 cases in which the human authored specifications were incorrect. That said, the specifications \ourtool produces are only as good as the tests on which it is evaluated. In 2 cases, \ourtool produces specifications that agree with \tshark, but \threedsymdiff detects, via symbolic differential testing, that the generated specification is semantically distinct from a human-authored EverParse sample---the \ourtool-produced specification does not enforce a constraint that it should. As such, we caution that \ourtool should not be used to blindly match the behavior of a legacy tool. Instead, we envision \ourtool and its symbolic tools to be used by humans to iteratively refine a natural language document into a formal specification, while also growing a carefully curated test suite.

A key enabler of our technique is the use of an effectively analyzable DSL, coupled with a verified code generator, as a medium of interaction between a user's informal intent and AI-generated output. In contrast, directly prompting AI agents to produce C code from informal specifications would leave open the question of analyzing ad hoc C code for safety and security, and with an unclear formal basis against which to assess code correctness. Further, targeting a DSL enables us to integrate powerful, fully automated tools like symbolic test-case generation and differential analysis that are usually intractable for large, general-purpose languages. We conjecture that future AI-assisted programming techniques might also benefit from the use of effectively analyzable DSLs as intermediate languages.

In summary, we make the following contributions:

\begin{enumerate}
\item An architecture for AI-assisted programming using DSLs coupled with symbolic analysis tools to refine informal user intent into formal specifications, grounded in the scenario of binary format parsers.

\item A new symbolic analysis and test generation tool for the 3D format language, integrated in \ourtool's intent refinement loop.

\item An evaluation of \ourtool on a suite of 20 binary format parsers specified in Internet standard RFCs, yielding safe and secure C code.
\end{enumerate}

%% file: body/2.background.tex
\section{Problem Formulation}
\label{sec:background}

Ramananandro et al.~\cite{everparse19usenix} present EverParse, a library of parser and serializer combinators in the \fstar programming language. They prove that every well-typed program assembled from their parser combinators produces a parser that is the inverse of the corresponding serializer, i.e.,  \lsf{forall s. parse (serialize s) == Some s} and \lsf{forall b v. parse b == Some v ==> serialize v == b}. Parsers for formats that satisfy this mutual inverse property are particularly relevant in security-critical settings. EverParse combinators are themselves embedded within a fragment of \fstar called Low*~\cite{lowstar}, which supports transpilation to C via a tool called Karamel.

Swamy et al.~\cite{everparse3d22pldi} present a DSL built on top of EverParse combinators called 3D, a language similar to C's language of type definitions, with typedefs, structures, and unions. 3D allows users to express a variety of ``tag-length-value" style formats, which are commonly used in many networking protocols and other variable-length formats. 

Internet RFCs often specify tag-length-value formats in natural language in an ad hoc way. For example, the TCP RFC 793 specifies the format of TCP options as follows:

\begin{footnotesize}
\begin{verbatim}
There are two cases for the format of an option:

Case 1:  A single octet of option-kind.
Case 2:  An octet of option-kind, an octet of
option-length, and the actual option-data octets.  
...
Currently defined options include
(kind indicated in octal):

    Kind     Length    Meaning
    ----     ------    -------
     0         -       End of option list.
     1         -       No-Operation.
     2         4       Maximum Segment Size.
...

 Maximum Segment Size

        +--------+--------+---------+--------+
        |00000010|00000100|   max seg size   |
        +--------+--------+---------+--------+
         Kind=2   Length=4

Maximum Segment Size Option Data:  16 bits
...
\end{verbatim}
\end{footnotesize}

To produce a parser for a TCP option in 3D, one starts by specifying the format declaratively. Here's one way to do it, defining an \ls`OPTION` as a structure with two fields: a byte field \ls`Kind` with a constraint that restricts its values to 0, 1, and 2; and a \ls`payload` field of type \ls$OPTION_OF_KIND(Kind)$, a type that \emph{depends on} the value of the \ls`Kind` field.

\begin{lstlisting}
typedef struct _OPTION {
    UINT8 Kind {
        Kind == 0x00 ||
        Kind == 0x01 ||
        Kind == 0x02
    };
    OPTION_OF_KIND(Kind) payload;
} OPTION;
\end{lstlisting}

The type \ls$OPTION_OF_KIND$, a \ls`casetype` in 3D, represents a form of \emph{union} type in C, where the parameter \ls`Kind` discriminates the case of the union. When \ls`Kind` is 0 or 1, the payload is empty, and when the \ls`Kind` is 2, the payload has type \ls$MAX_SEG_SIZE$.

\begin{lstlisting}
casetype _OPTION_OF_KIND(UINT8 Kind) {
    switch (Kind) {
        case 0x00: unit case0; /*unit: empty payload*/
        case 0x01: unit case1; /*unit: empty payload*/
        case 0x02: MAX_SEG_SIZE case2;
    }
} OPTION_OF_KIND;
\end{lstlisting}

Finally, the type \ls$MAX_SEG_SIZE$ is a structure, with one byte for the \ls`Length` and 2-bytes for an unsigned 16-bit big-endian integer for the \ls`MaxSegSize`.

\begin{lstlisting}
typedef struct _MAX_SEG_SIZE {
   UINT8 Length;
   UINT16BE MaxSegSize;
} MAX_SEG_SIZE;
\end{lstlisting}

From this specification, EverParse (in its simplest mode) generates a C program with the following signature, a function \ls`CheckOption` which, when called with a byte buffer \ls`Input` containing at least \ls`Length` bytes, checks that \ls`Input` contains a valid representation of \ls`OPTION`, returning an error code recording success, or details about where and why validation failed.

\begin{lstlisting}[language=C]
EVERPARSE_ERROR_CODE CheckOption(
    UINT8* Input,
    UINT64 Length)
\end{lstlisting}

A 3D user turning an ad hoc description from an RFC to a specification must convince themselves that they have captured the intent of the RFC---a process that typically involves careful review combined with testing. While 3D was designed to be used by C programmers and benefits from its resemblance to C, the constructs it offers, including type dependency, value constraints, parameterization, case analysis, etc. take effort to learn and use correctly. Further, while EverParse guarantees that the generated C code is memory safe, free from bugs that trigger undefined behaviors, and faithfully parses exactly the specified format, the source specification is still subject to audit. For example, one could easily have specified \ls`UINT16 MaxSegSize`, forgetting a convention that networking protocols like TCP typically use big-endian integers---users need assistance in specification testing and validation.

Swamy et al.~\cite{everparse3d22pldi} report that 3D has been used to specify a suite of networking protocols used in production software at Microsoft, including in the kernel of the Windows 11 release. They report using 3D to specify ``137 structs, 22 casetypes, and 30 enum type definitions" in around 5,000 lines of 3D specifications, stating that ``describing those message formats required careful specification  engineering and discovery" over a period of 18 months. With \ourtool, we seek to lower this overhead, 
making 3D accessible to non-experts by directly synthesizing specifications from informal intent, and to offer systematic testing tools to assist with validation.

%% file: body/3.3DGen.tex
\section{The \ourtool Approach}

\input{body/adsl}

\input{body/3DAgent}

\input{body/testgeneration}

%% file: body/adsl.tex
In this Section, we make precise the workflow in Figure~\ref{fig:mainworkflow}, showing how
we derive a 3D specification from RFCs and tests.
We start by describing the algorithm abstractly, parameterized by several non-deterministic choices, AI-based components, and a test generator for 3D. We then describe an {\it agent-based} implementation of the AI-based components, and \symbTestGen, a new symbolic test generator for 3D.

\subsection{\ADSLProgGen: An Abstract Algorithm}
\label{subsec:3dcodegen}

We assume the 3D DSL is equipped with the following functions: 
\begin{itemize}
    \item $\SC$: A syntax and type checker function, given a specification $p$  checks for syntax as well as type constraints imposed by the 3D language. 
    \item $\Ex$: An execution function; for any 3D specification $p$ that satisfies $\SC(p)$, given a packet $i$, $\Ex(p, i)$ returns true iff the specification $p$ accepts the packet $i$. Concretely, we use EverParse to compile $p$ to C code and execute it on $i$.
  
    \item $\DslDoc$: Natural language documentation about the 3D language and examples, provided as a manual. 
\end{itemize}

\newcommand{\Impl}{\textit{LblImpl}}
\newcommand\RFC{\ensuremath{\mbox{\textit{RFC}}}\xspace}
Algorithm~\ref{alg:3dcodegen} takes as input an \RFC document, function $\Impl$ that is used to classify packets, as well as a (possibly empty) seed sets of positive and negative packets, $\posSeedTests$ and $\negSeedTests$. 
The desired 3D specification $p$ should accept the positive packets $\posSeedTests$ and reject $\negSeedTests$. 
The algorithm returns a set of possible candidate 3D specifications $\candProgs$, along with an augmented set of positive ($\posTests$) and negative ($\negTests$) packets, generated by $\symbTestGen$ and labeled using $\Impl$, ensuring that every specification in $\candProgs$ is consistent with the augmented set of packet inputs. 
Formally, $\posSeedTests \subseteq \posTests$, $\negSeedTests \subseteq \negTests$, and for each $p \in \candProgs$, $\Ex(p, i^+) = \text{true}$ for each $i^+ \in \posTests$, and $\Ex(p, i^-) = \text{false}$ for each $i^- \in \negTests$.

The algorithm iterates non-deterministically, accumulating state which records all relevant information to be fed to an LLM in $st$, initialized to the $\DslDoc$, the \RFC, and the seed tests.
At each iteration, it performs one of the following two actions non-deterministically:
(i) augment $\candProgs$ with a new well-formed 3D specification $p$ by querying an LLM (lines~\ref{line:block1start}--\ref{line:block1end}); 
or, (ii) augment the labeled packets in $\posTests$ and $\negTests$ using $\symbTestGen$ and $\labelTest$ (lines~\ref{line:block2start}--\ref{line:block2end}).
Finally, at lines~\ref{line:prunestart}--\ref{line:pruneend}, candidates in $\candProgs$ that are not consistent with the labeled packets are pruned, and any failing candidate/test pairs are added to the state (line~\ref{line:failstest}). 

To generate a candidate program $p$, at line~\ref{line:queryllm} we $\queryLLM$ with the accumulated state---this step is implemented using agents, as described in Section~\ref{subsec:3dagent}.
If $p$ fails the $\SC$, we update the state with the error, and retry.

The 3D symbolic test generator $\symbTestGen$ takes as input a set of well-formed candidate programs that satisfy the current $\posTests \cup \negTests$, and outputs new (unlabeled) packets by symbolically analyzing the programs in $\candProgs$; we describe the precise implementation in Section~\ref{subsec:3dtestgen}.
We then use $\Impl$ to label each packet as positive or negative---concretely, we use \tshark as an implementation of $\Impl$. Details of $\labelTest$ is present in Section~\ref{subsec:labeltests}.

\newcommand\aset[1]{\{#1\}}
\newcommand\getsplus{~+\!\!=~} 

\begin{algorithm}
\caption{\ADSLProgGen{} Algorithm}
\label{alg:3dcodegen}
\scriptsize{
\begin{algorithmic}[1]
\REQUIRE $\Req, \Impl, \posSeedTests, \negSeedTests$
\ENSURE  $\candProgs, \posTests, \negTests$ 
\STATE $(\posTests, \negTests) \gets (\posSeedTests, \negSeedTests)$
\STATE $st \gets \aset{\DslDoc, \Req, \posTests,  \negTests}$ \label{line:initprompt}
\STATE $\candProgs \gets \aset{}$
\FOR{$*$}
     \IF {$*$} \label{line:block1start}
        \STATE $p \gets \queryLLM(st)$ \label{line:queryllm}
        \STATE $se \gets \SC(p)$ 
        \IF{$se \neq \text{SUCCESS}$}
            \STATE $st \gets st \cup \aset{(p,se)}$ \label{line:repairsyntaxprompt}
            \STATE \textbf{continue}
        \ENDIF
        \STATE $\candProgs \gets \candProgs \cup \{p\}$ 
    \ENDIF \label{line:block1end}
    \IF {$*$} \label{line:block2start}
        \STATE $I' \gets \symbTestGen(\candProgs, \posTests \cup \negTests)$ \label{line:testgen}
         \STATE $(\posTests, \negTests) \gets (\posTests, \negTests) \cup  
    \labelTest(I', \Impl)$ \label{line:label}
    \ENDIF \label{line:block2end}
    \FOR {$q \in \candProgs$} \label{line:prunestart}
        \FORALL{$i \in \posTests \cup \negTests$} 
            \IF{$ 
             \bigvee \left( \begin{array}{cc}
              & i \in \posTests \land \neg \Ex(q, i) \\
              & i \in \negTests \land \Ex(q, i) 
             \end{array}  \right)
             $} \label{line:failstest}  
                \STATE $st \gets st \cup \aset{(q, i)}$ \label{line:repairfailprompt}
                \STATE $\candProgs \gets \candProgs \setminus \{q\}$
                \STATE \textbf{break}
            \ENDIF
        \ENDFOR
    \ENDFOR \label{line:pruneend}
\ENDFOR
\RETURN $\candProgs, \posTests, \negTests$
\end{algorithmic}
}
\end{algorithm}

%% file: body/3DAgent.tex
\subsection{Agent Based Implementation}
\label{subsec:3dagent}

Constructing a prompt for an LLM from the diverse information in Algorithm~\ref{alg:3dcodegen}'s accumulated state is non-trivial. It involves choosing relevant sections of natural language documentations from several pages of $RFC$, $\DslDoc$, relevant examples, failing tests to focus on, etc. Composing a single monolithic prompt with all relevant context needed to solve a task is often impossible, given the restricted token context-window for LLMs.

Instead, research shows that LLM-based \emph{agents} significantly extend the capabilities of standalone LLMs by equipping them with the abilities needed to solve tasks in a self-directed fashion, such as long-term planning, reasoning\cite{yao2022react}, conversing with other LLMs, using tools, and retrieving information critical to task resolution\cite{gao2023retrieval}. Agents demonstrate improved performance and generalization of task resolution abilities for a number of increasingly complex and real-world tasks \cite{xi2023rise, wang2023survey}.  Furthermore, orchestrating multiple agents that are instructed to cooperate together, can scale up the capabilities of a single agent by decomposing tasks, improving factuality and reasoning \cite{du2023improving}, and validation \cite{qian2023communicative}.

 Motivated by these findings,  we design an agent system based on the AutoGen\cite{wu2023autogen} multi-agent framework. AutoGen allows the instantiation of multiple agents, each unique in their task description, access to tools and  inputs, and instructed to cooperate together to achieve a solution.  We choose to use a multi-agent framework over a single agent, to decompose tasks and reduce overall input in the context window, shielding other agents from unrelated intermediate reasoning steps involved in distinct task refinement loops. In the AutoGen multi-agent setup agents converse in a \emph{group chat} setting, critiquing and reflecting on task progress based on conversation history, and adapting from feedback. AutoGen provides the multi-agent conversation framework as a high level abstraction, requiring only meta prompts and tool customization from the user. The agent framework designed for \ourtool is not reliant on one particular LLM, however we use GPT4-32k across all experiments. The \ourtool multi-agent framework is implemented concretely as three distinct agents:

\begin{enumerate}
 \item \textbf{Planner Agent:} the planner agent is a tool-backed agent that orchestrates the multi-agent conversation. In AutoGen it is instantiated as the \emph{group chat manager}. It has access to the meta task prompt, descriptions of the other two agents, and the ability to invoke tools $\SC$ and $\Ex$, and communicate the results to the other agents. 
    \item \textbf{3D Developer Agent:} the 3D developer agent is tasked with generating 3D code based on  instructions communicated from the other two agents. This agent has access to the full 3D language manual $\DslDoc$, a set of task examples, and high level tips about generating syntactically correct 3D code.  
    \item \textbf{Domain Expert Agent:} the domain expert agent is tasked with communicating specifications to the 3D Developer Agent, and critiquing generated 3D code. It has access to all domain-relevant documents needed to solve the problem. In this work, the domain expert agent has access to an RFC, the network protocol specification document needed to solve the task, and examples of the task. 
\end{enumerate}

All three agents communicate via an inter-agent group chat, until one of two termination conditions are met: either the output of $\Ex$ indicates that the generated 3D specification passes on the test set, or the maximum number of iterations, as set by the user, have been completed. The control flow of task resolution follows the paradigm provided in AutoGen, and is entirely conversation-driven, i.e. the participating agents’ decisions on which agents to send messages to and the procedure of computation are functions of the inter-agent conversation. An example of control flow is as follows: 1) the Domain Expert agent communicates the relevant parts of the RFC specification to the 3D Developer Agent 2) the 3D Agent generates a candidate specification 3) the Planner Agent makes a call to $\SC$ and communicates the result to the group 4) the 3D Agent reflects on  a syntax error reported by the Planner and refines the specification. One example of the control flow is later shown in Figure~\ref{fig:vxlan}.

%% file: body/testgeneration.tex
\subsection{\threedsymdiff: Symbolic Test Case Generation}
\label{subsec:3dtestgen}
In this section, we discuss our implementation of the \symbTestGen{} sub-routine of Algorithm 1. Our test (packet) generator, called \threedsymdiff, is implemented as an extension of the EverParse toolchain, and is grounded in the formal semantics of 3D. \threedsymdiff encodes 3D programs into the SMT-LIB version 2 language (a.k.a. SMT2)~\cite{SMTLIB}, relying on SMT-solver Z3~\cite{de2008z3} to produce test cases. Given a 3D program $p$, to obtain positive (resp. negative) test cases, \threedsymdiff encodes the semantics of $p$ to Z3 and asks for models of the existential predicate: "does there exist a sequence of bytes that makes $p$ succeed (resp. fail)". Z3 returns with one of the following answers:
\begin{itemize}
    \item SAT: Z3 finds a model to satisfy predicate, including a concrete sequence of bytes that makes the parser succeed (resp. fail)
    \item UNSAT: the predicate is \emph{unsatisfiable}, which means that $p$ always fails (resp. succeeds)
    \item UNKNOWN: Z3 times out. While there may be multiple causes to Z3 timeout; in our case, this happens rarely.  
\end{itemize}

Further, given two 3D programs $p_1$ and $p_2$, \threedsymdiff can also produce differential test cases by asking Z3 to find a sequence of bytes that satisfy $p_1$ but not $p_2$.
If Z3 returns UNSAT, then there are no such test cases: every packet satisfying $p_1$ also satisfies $p_2$. Separately, we can ask Z3 the same satisfiability question but with $p_1$ and $p_2$ swapped. If Z3 returns UNSAT for both ways, then the 3D programs $p_1$ and $p_2$ accept and reject the exact same packets: they are semantically equivalent. We do not bound the size of packets. %

The semantics of a 3D program is represented by a pure \fstar function whose (simplified) signature is a value of type \lsf{parser t}, a function of the form 
\lsf{(input:seq byte) -> option (t * nat)} which when applied to an input sequence of bytes may fail returning \lsf{None}, or succeed with \lsf{Some (v, n)}, where \lsf{v} is the parsed value (of type \lsf{t}) and \lsf{n} is the number of bytes of the input that were consumed.

As described in \S\ref{sec:background}, EverParse provides \emph{combinators}, library functions that  allow combining simpler parsers to build more complex ones in a correct-by-construction way. For example, the following 3D program defines a \lss{message} data format specification as a structure of two unsigned 8-bit integer fields, \lss{first} and \lss{second}, where \lss{first} has a constraint on its value:

\begin{lstlisting}[language=3D]
typedef struct _message {
  UINT8 first { first > 42 };
  UINT8 second;
} message;
\end{lstlisting}

\noindent The semantics of that 3D program is modeled in \fstar by the following \lsf{message} parser \lsf{parse_message} as an application of the \lsf{parse_pair} combinator:

\begin{lstlisting}[language=fstar]
let parse_message =
parse_pair (parse_refine parse_uint8 (fun first -> first > 42))
            parse_uint8
\end{lstlisting}

\noindent where \lsf{parse_pair}, defined in EverParse, has a higher-order type \lsf{parser t1 -> parser t2 -> parser (t1 * t2)}. 
Operationally, the code above first checks that the input sequence contains at least one byte; then parses the first byte using \lsf{parse_uint8} and reads it into the variable \lsf{first}; then advances the position in the input by one byte; then checks that \ls`first < 42`; then tries to read and return the next byte. If any of the checks fail, the parser returns \ls`None`. Otherwise, it returns \ls`Some ((first, second), 2)`, a pair containing the two bytes that were read and \lsf{2} to indicate that two bytes were consumed.

We would like to encode \lsf{parse_message} to Z3, but SMT2 does not 
directly support higher-order functions like \lsf{parse_pair}. So, at the heart of \threedsymdiff is a specialization pass over the 3D parser-combinator AST to turn it into a first-order program.

For starters, the SMT2 semantics of a 3D program is given in the context of an uninterpreted function \lss{Input} representing the input byte sequence on which the parser operates. The assertion below constrains Z3 to pick models for the \lss{Input} array where every element is a byte in the range $[0, 256)$.
\begin{lstlisting}[language=SMT2]
(declare-fun Input (Int) Int) 
(assert (forall ((i Int))
                (and (<= 0 (Input i)) (< (Input i) 256))))
\end{lstlisting}

Next, we encode a 3D program 
as an SMT2 \lss{State} transforming function, where the \lss{State} records (among other things) the remaining size of the \lss{Input} byte sequence (\lss{remaining-input-size}); the number of bytes read so far (\lss{current-pos}); whether or not the parser has failed (\lss{has-failed}); and the value they return (\lss{return-value}). 

We start by showing a simplified encoding of \lss{parse_uint8}:

\begin{lstlisting}[language=SMT2]
(define-fun parse-uint8 ((s0 State)) State
  (if (and (not (has-failed s0))
           (> (remaining-input-size s0) 0))
      (success-state 
        (Input (current-pos s0)) ;; return value.
        (incr (current-pos s0))  ;; new position.
        (decr (remaining-input-size s0))) ;; new remaining size.
      (fail-state s0)))
\end{lstlisting}

Next, we show the encoding of the \lsf{parse_message} parser, where \threedsymdiff has inlined the \lsf{parse_pair} and \lsf{parse_refine} higher-order combinators from the 3D AST. This representation of the \lsf{parse-message}  is adequate for test-case generation. 

\begin{lstlisting}[language=SMT2]
(define-fun parse-message ((s0 State)) State
  (let ((s1 (parse-uint8 s0)))
    (if (has-failed s1) s1
      (if (> (return-value s1) 42)
          (parse-uint8 s1)
          (fail-state s1)))))
\end{lstlisting}

Given a query such as the following, which constrains the initial state \lss{init} and 
asserts that \lsf{message} parser \lss{parse-message} does not fail when applied to \lss{init}, Z3 can produce models for the \lss{Input} variable and its length in bytes, yielding a positive test case; 
to generate a negative test case, we would assert instead that \lss{(has-failed (parse-message init))}. Since the input byte sequence can be arbitrarily long, we do not constrain its initial size.

\begin{lstlisting}[language=SMT2]
(declare-fun init () State) ;; initial state.
(assert (and (not (has-failed init))
             (= 0 (current-pos init))))
(assert (not (has-failed (parse-message init))))
(check-sat)
(eval (remaining-input-size init)) ;; input size from model.
(eval (Input 0)) ;; retrieve first input byte.
\end{lstlisting}

If Z3 returns a model, then \threedsymdiff can iteratively query Z3 for further distinct models with additional appropriate assertions to avoid duplicates. However, with this encoding, Z3 generates models that may not cover all possible branches. Our implementation uses a more sophisticated encoding to track which branches of a specification are covered by a given test case, allowing us to systematically generate tests that cover all branches up to a user-provided branch depth. We provide more details in the Appendix.

Leveraging the existing semantics of 3D and its compact, structured language of parser combinators, our implementation of \threedsymdiff took less than 3~person-weeks and around 2000~lines of \fstar, OCaml, and SMT2.

%% file: body/4.Dataset.tex
\section{Experimental Setup}
\label{sec:dataset}

\subsection{Network Protocols}
We evaluate \ourtool on 20 network protocols specified in IETF standards. Table \ref{tab:protocols} lists each protocol and a short description and the specific RFC number. We include the length of the RFC in pages, as a rough measure of complexity, though RFCs contain a lot of information beyond the description of the format---the number in parenthesis, when present, shows the number of pages in the RFC concerned with header formats.

\begin{table}[h!]
\resizebox{\columnwidth}{!}{
\begin{tabular}{@{}rl|ccl@{}}
\toprule
\multicolumn{1}{l}{\textbf{\#}} & \textbf{Protocol} & \begin{tabular}{c} \textbf{RFC} \\ \small{(Version)}
\end{tabular}  & \begin{tabular}{c} \textbf{Length} \\ \small{(Pages)}
\end{tabular} & \textbf{Description} \\ \midrule
1 & UDP* & 768 & 3 &  User Datagram Protocol \\
2 & ICMPv4 * & 792 & 21 &  Internet Control Message Protocol \\
3 & VXLAN* & 7348 & 22 & Virtual eXtensible Local Area Network \\
4 & IPV6* & 2460 & 39 (24) & Internet Protocol version 6 \\
5 & IPV4* & 791 & 45 (12)  &   Internet Protocol version 4 \\
6 & TCP* & 793 & 85 (10) &   Transmission Control Protocol \\
7 & Ethernet* & 7348 & 22 & Ethernet II Frames in VXLAN\\ \hline
8 & GRE & 2784 & 9 &  Generic Routing Encapsulation \\
9 & IGMPv2 & 2236 &  24&  Internet Group Managment Protocol \\
10 & DHCP & 2131 & 45 (4) & Dynamic Host Configuration Protocol\\
11 & DCCP & 4340 & 129 (14) &  Datagram Congestion Control Protocol \\
12 & ARP & 826 & 10 &  Address Resolution Protocol \\
13 & NTP & 5905 & 110 (4) &  Network Time Protocol  \\
14 &  NBNS & 1002 &84 (6)&  NetBIOS Name Service  \\
15 & NSH & 8300 & 40 (8) &  Network Service Header \\
 
16 & TFTP & 1350 & 11 &   Trivial File Transfer Protocol\\
17 & RTP & 3550 & 104 (3) &  Transport Protocol for Real-Time Applications \\
18 & PPP & 1661 &  52 (11) &  Point-to-Point Protocol \\ 
 
19 & TPKT & 2126 & 25 &  ISO Transport Service on top of TCP\\
20 & OSPF & 5340 & 94 (13)  &  Internet Official Protocol Standards  \\
\bottomrule \\
\end{tabular}
}
\caption{Dataset of protocols and corresponding RFCs. * denotes protocols for which there is a human written 3D specification. Page numbers in ( ) indicate the length of the extracted RFC.}
\label{tab:protocols}
\end{table}

\subsection{Generating Specifications with \ourtool}
To evaluate \ourtool's ability to translate natural language specifications into 3D format specifications, we use it to generate 5 candidate specifications for the protocols in Table~\ref{tab:protocols}. We deem a
generation successful if the produced specification correctly classifies a test set of generated packets labeled by Wireshark. We report a \emph{pass@5} metric, which counts the number of successful generations
out of 5 runs. To evaluate how well agents in the \ourtool multi-agent framework are able to understand the natural language specifications contained in the network protocol RFC, as well as its ability to learn 3D syntax, we explore the number of refinement loops needed to generate 1) a syntax- and type-correct solution as determined by $\SC$ and 2) a semantically correct solution with respect to the test set, as determined by $\Ex$. In addition, we highlight common mistakes made by the agents, as well as several instances where the agent is able to learn constraints from the RFC that are not enforced by the Wireshark. 

For each protocol we instantiate \ourtool, using GPT-4-32k as the underlying LLM, at temperature 1.0. We set the number of specifications to generate to 5 and the max number of syntax or packet refinements to 15 per attempt. If the agents are able to produce a specification that passes before 15 refinements, the agent loop is terminated. We observe that allowing more refinement loops within the \ourtool agent conversation flow does not always lead to a successful attempt at generating a specification. On the other hand, reducing the number of refinement loops often does not give the agents enough attempts at solving the problem. This is especially true when the protocol is more complex, requiring the agents to produce a longer specification, for example in the case of ICMP, where there are 8 distinct message types for which the agent must generate a 3D specification.

For each protocol, we download the RFC from the IETF Data Tracker\footnote{https://datatracker.ietf.org} as a text file to provide as input in the 3D Developer agent prompt. In some cases, the full RFC is prohibitively long, and would exhaust the GPT-4-32k token window. Since we are only interested in the part of the specification related to the header data format, in such cases we manually extract the pages of the RFC related to the data format specification, (usually labeled in a section called ``Header Specification"). The length of the extract pages is denoted in ( ) in Table~\ref{tab:protocols}.In the future, we plan to explore building a retrieval tool specific to RFC  extraction.

\subsection{Generating a Labeled Test Set for each Protocol}
\label{subsec:labeltests}

We build a test suite consisting of a mixture of real-world packet captures and synthetically generated tests by \symbTestGen, where we use Wireshark to decide if the packet should be considered valid or not. For each protocol, we collected a small number of real world packets from various sources on the Internet and retain only those that Wireshark considers valid---we discard the negative cases, since they are sometimes arbitrarily malformed. For synthetic tests, we run one iteration of the \ourtool loop seeded with the real-world packets. This produces a single candidate specification on which we run \symbTestGen to generate 200 test cases. We configure \symbTestGen to fully explore a trace with 100 branch points, and observed that while many examples have a large number of branches (e.g., ICMP has 82 branches), none of them have more than 100 branches. We then collect at least two examples from each branch point, until we reach 200 examples. This gives us some confidence that the generated test suite is diverse, though its completeness is limited by the quality of the specification on which  \symbTestGen is executed. We then use Wireshark to label the tests, obtaining both positive and negative test cases.

Using Wireshark to label packets comes with its own challenges. For starters, Wireshark is a protocol analyzer typically used for diagnostics and experimentation and, by design, does not always enforce all constraints when validating a packet. Wireshark does not implement a dissector for a single RFC, but rather for a family of RFCs.  Thus, a dissector may be more permissive compared to a given RFC, perhaps because a related RFC mandates such a behavior---our experiments in Section~\ref{result} uncover many such unenforced constraints. To label a test case produced by \symbTestGen, we rely on a Wireshark feature called \emph{Export PDU}, which allows validating a given packet header without encapsulating it within outer protocol headers. Two exceptions were Ethernet, which does not require outer encapsulation; and TFTP which is not supported by Export PDU. We wrapped TFTP headers generated by \symbTestGen with a dummy UDP header. For Ethernet, IPv4, IPv6, VXLan, TCP, and UDP, we also had to generate dummy payloads to prevent Wireshark from raising trivial errors. We also had to disable checksums, since this is not enforced at the level of the formats. As such, using Wireshark as a labeler for \symbTestGen outputs involved a non-trivial effort.

\subsection{Handwritten 3D Specifications}
The EverParse GitHub\footnote{https://github.com/project-everest/everparse} repository contains 3D specifications for seven network protocols, written by the authors of EverParse. Protocols with an existing handwritten 3D specification are marked with a \textbf{*} in Table \ref{tab:protocols}. In Section~\ref{sec:rq3} we use \symbTestGen{} to compare the specifications generated by \ourtool to the seven handwritten specifications. We report if any of the specifications produced by \ourtool are semantically equivalent to the handwritten specifications, and otherwise use the generated set of tests to identify the cause(s) of differences.

%% file: body/5.results.tex
\section{Results}
\label{result}

\input{body/rqs/rq1}
\input{body/rqs/rq2}

\input{body/rqs/rq3}

%% file: body/rqs/rq1.tex
\subsection{Capabilities of \ourtool}
\label{sec:rq1}

In this section, we present experimental evidence to answer the following central question underpinning our work: 

\noindent\textbf{RQ1:} \rqa

Table \ref{tab:rq1} details results of \ourtool on our dataset of 20 network protocols, We report
the pass@5 metric, the average number of syntax refinement, and packet feedback refinement loops across all attempts.

At a first glance, \ourtool is able to generate a passing specification for 9/20 protocols, with a pass@5 $= 45\%$. For two protocols, such as UDP and IGMP, \ourtool generates a passing specification in all 5 attempts, requiring an average of 0.3 syntax refinements and 0 packet refinements for UDP, and 3.8 and 0 for IGMP respectively. For the other seven, IPV4, DHCP, DCCP, ARP, NTP, NBNS, and NSH, \ourtool  can generate at least one passing specification, but is not successful in all 5 attempts.  

For the remaining 11 protocols, \ourtool is  unable to generate a specification that passes on the test set labeled by Wireshark.  We investigate the cause of the errors and find in all cases that \ourtool generates a specification that is consistent with the header format described in the RFC, but that Wireshark labels packets as valid despite there being constraint violations, i.e., Wireshark does not fully enforce the RFC. In some cases, Wireshark emits a warning about these violations, which can be filtered on a case-by-case basis. However, this is not a straightforward task as some warnings do not impact the data format specification, e.g.,  Wireshark emits a warning when a packet indicates the TCP connection is reset. Besides, in most cases a warning is not reported. 

Using the RFC as a guide, two authors manually identified tests that Wireshark labels too permissively and corrected the labels. Using these corrected labels, we checked if any of the \ourtool generated specifications already pass on this modified test set, or else restart the \ourtool loop with the corrected tests. Protocols for which \ourtool is able to generate a specification that passes on the corrected test set are denoted as (1*) in Table \ref{tab:rq1}---\ourtool is able to generate at least one candidate specification that is consistent with the labels in \emph{all} cases.

\subsubsection{Labeler and RFC disagreement}

We look more closely at examples where Wireshark is too permissive: Table~\ref{tab:disagreement} details the cause(s) of disagreement between the Wireshark labels and the constraints specified by the RFC in each case. 

For example, for RTP, RFC 3550\footnote{https://www.ietf.org/rfc/rfc3550.txt} states that the \lstt{version} field is 2 bits and \emph{``The version defined by this specification is two (2)"}. Although the RFC does not indicate a strong constraint that the version field must be set to 2, a large number of negative packets in the RTP test set include version numbers other than 2. Wireshark labels these packets as acceptable, however the \ourtool agent consistently generates a specification with \ls`UINT16BE Version:2 {Version == 2}`. In this case, the agent is unable to learn from feedback about why a test with a different version value should pass, because Wireshark does not provide any warning. Enforcing this constraint by changing the Wireshark label to negative allows the specification to pass all tests. 

Similarly for TFTP, RFC 1350\footnote{https://datatracker.ietf.org/doc/html/rfc1350} states that a TFTP header contains a 2 byte \lstt{opcode} field and enumerates 5 possible opcode values. \ourtool always produces a specification constraining the value of the opcode field to one of the 5 values in the RFC. Wireshark does not label packets with other opcode values as malformed. When we manually label tests with incorrect opcode fields as negative,  the specification generated by \ourtool  passes on the test set.

In the case of GRE, RFC 2784\footnote{https://datatracker.ietf.org/doc/html/rfc2784} states \emph{``The Version Number field MUST contain the value zero"}. However, Wireshark does not enforce version constraints on GRE, likely because the for the PPTP variant of GRE, the version field is set to 1, and Wireshark uses the same dissector for all GRE versions.

In the case of IPV6, the generated specification consistently fails to reject packets labeled as malformed by Wireshark. The IPV6 \lstt{NextHeader} field indicates the value of the encapsulated packet, however 
the RFC focuses only on the constraints of a single layer, not of any encapsulated packets. In contrast, Wireshark validates packet contents across layers and rejects packets that don't encapsulate the next layer correctly.

\begin{figure}
    \centering
    \includegraphics[width=\columnwidth]{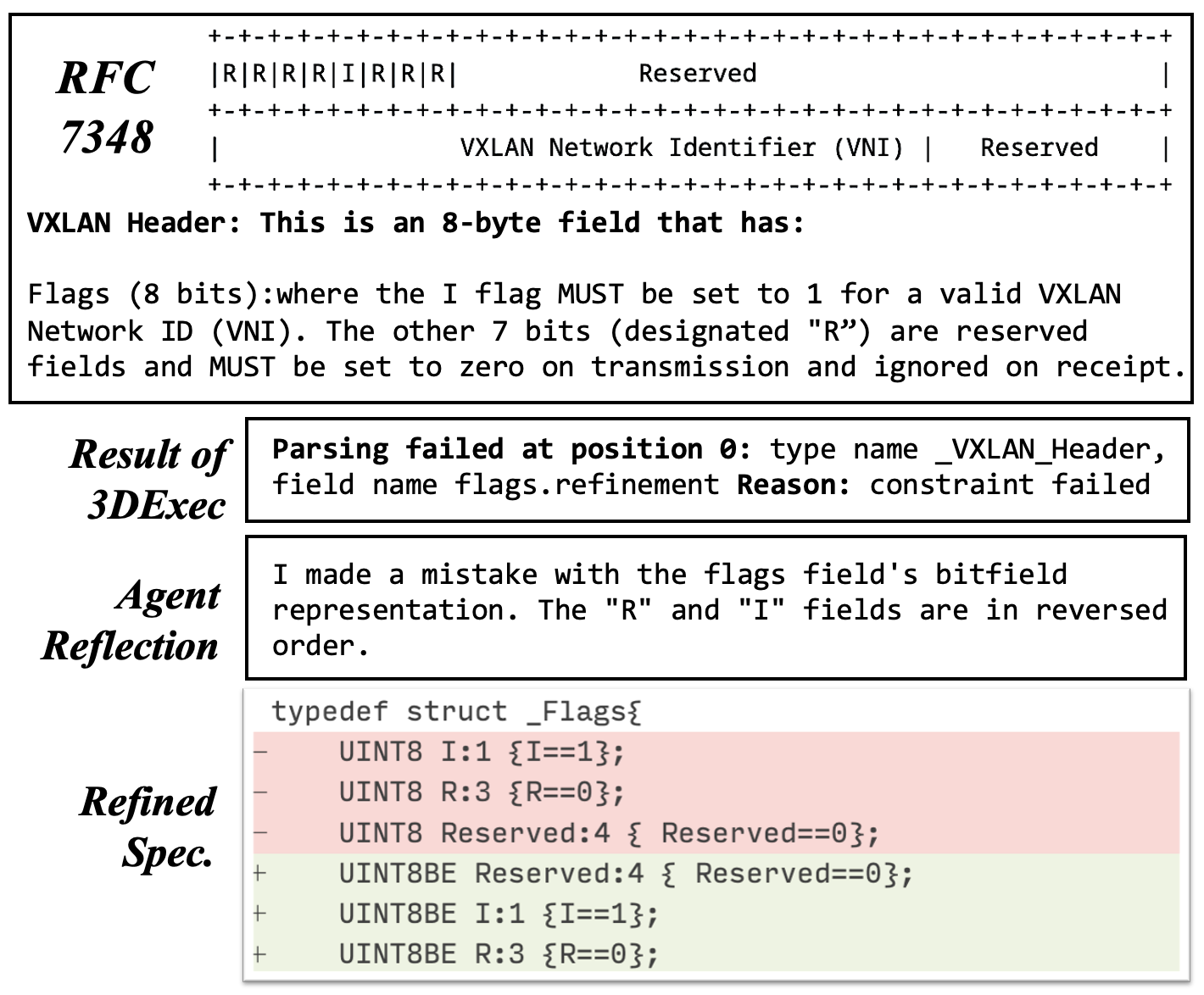}
    \caption{ Example of a packet refinement loop by the \ourtool agent for VXLAN RFC 7348 }
    \label{fig:vxlan}
\end{figure}

\subsubsection{Agent Mistakes}
From Table~\ref{tab:rq1}, we observe that the agents frequently make syntax mistakes, despite having access to the language manual for the 3D language. However, given that 3D is not yet a widely used DSL, syntax mistakes from a model like GPT-4 are anticipated. We observe that the agents struggle most with using correct 3D bitfield notation. Using only the primitives that 3D supports (\ls`UINT8`, \ls`UINT16`, \ls`UINT32`, \ls`UINT64` and their BE counterparts) and refraining from using reserved keywords like `type' as identifier names also require refinement steps. On the other hand, the agents are able to easily learn some other 3D specific constructs, such as the \ls`consume-all` notation. We also observe that semantic mistakes, i.e., incorrect specifications, can stem from the agent's difficulty in understanding the natural language in the RFC document. Figure \ref{fig:vxlan} shows one such example for VXLAN. The RFC describes the fields in the header in natural language, and also provides an ASCII diagram. However, the natural language description of the fields do not indicate the order in which the different values should occur in the VXLAN header, and without the ASCII diagram a reader would not be able to correctly interpret the RFC. The agent first produces a specification  which is then executed using $\Ex$, resulting in a parsing error on the flags field, for one of the tests in the test set. Then the agent reflects on the result of $\Ex$ and determines that the parsing failed due to the mis-ordering of the \lstt{Reserved} and \lstt{I} fields. It then refined the specification by flipping the order of the fields (4) and produces a candidate that passes on all tests. Interestingly, the language in the RFC first describes the \lstt{I} flag and then says specifies the values of the other 7 bits, without indicating that there are 4 reserved bits, followed by an I bit, followed by the remaining 3 bits. Ambiguities in the RFC language may cause the agent to misinterpret the true intent.

\RS{1}{\ourtool is able to generate syntactically correct 3D code, and learn from mistakes to refine specifications. 
Even in the presence of a noisy labeler and non-exhaustive tests, 3DGen enables users to leverage the generated specifications to align the test set with the RFC, yielding specifications that pass all aligned tests for all 20 network protocols.
 }

\begin{table}[h]
\centering
\resizebox{0.8\columnwidth}{!}{%
\begin{tabular}{@{}l|ccc@{}}
\toprule
\textbf{Protocol} & \textbf{\begin{tabular}[c]{@{}c@{}}Accepted\\ (x/5)\end{tabular}} & \textbf{\begin{tabular}[c]{@{}l@{}}Avg. Syntax\\ Refinements\end{tabular}} & \textbf{\begin{tabular}[c]{@{}l@{}}Avg. Packet\\ Refinements\end{tabular}} \\ \midrule
UDP & 5 & 0.3 & 0 \\ 
ICMP & 0 (1$^*$) & 7.5 & 6 \\ 
VXLAN & 0 (2$^*$) & 7.6 & 7.4\\ 
IPV6 & 0 (1$^*$) & 7.8 & 2.0 \\ 
IPV4 & 2 & 4 & 11 \\
Ethernet & 0 (1$*$)  & 11.4 & 4.6 \\
TCP & 0 (2$*$)  & 10.2 & 2.5 \\
GRE & 0 (1$*$) & 10.4 & 4.6 \\
DHCP & 2 & 8.2 & 0 \\
DCCP & 1 & 14.25 & 0.75 \\
TPKT & 0 (1$^*$) & 5.0 & 6.6 \\
ARP & 3 & 4.8 & 1.8 \\
NTP & 3 & 7.4 & 3 \\
NBNS & 1  & 4.6 & 4 \\
IGMP & 5 & 3.8 & 0 \\
NSH & 1 & 11.0 & 2.0 \\
TFTP & 0 (1$^*$) & 11 & 1 \\
RTP & 0 (1$^*$) & 3 & 12 \\
PPP & 0 (1$^*$) & 7.6 & 5 \\
OSPFv3 & 0 (1$^*$) & 13.6 & 2 \\ \midrule
 & pass@5: (45\%)  & pass*@5: (100\%) & \\ \bottomrule
\end{tabular}%
}
\caption{ Results of \ourtool for 20 network protocols. * denotes protocols for which the test labels were adjusted to be consistent with the RFC.}
\label{tab:rq1}
\end{table}

\begin{table}[]
\centering
\resizebox{\columnwidth}{!}{%
\begin{tabular}{@{}lll@{}}
\toprule
\textbf{Protocol} & \textbf{Detected RFC vs Wireshark Disagreement} & \begin{tabular}{l} Wireshark\\ Message 
\end{tabular} \\ \midrule
ICMP & Header length constraints & None \\ \midrule
Ethernet & \lstt{Ethertype} payload length & None \\ \midrule
VXLAN & \begin{tabular}[c]{@{}l@{}}\lstt{Reserved} bits must be 0\\ \lstt{I flag} must be 1 \\ Header must be 8 bytes\end{tabular} & \begin{tabular}[c]{@{}l@{}}None \\ None \\  None \end{tabular} \\ \midrule
IPV6 & \lstt{Payload Length} exceeds framing length    &  Warning   \\ \midrule
GRE & \begin{tabular}[c]{@{}l@{}}\lstt{Reserved0} must be zero \\ \lstt{Version} number must be zero  \end{tabular}& \begin{tabular}[c]{@{}l@{}} None \\ None  \end{tabular}  \\ \midrule
TFTP & \lstt{Opcode} fields must be between 0-5 & None\\ \midrule
TCP & \begin{tabular}[c]{@{}l@{}}\lstt{Window} fields must not be zero \\ \lstt{ACK number} must be consistent with \lstt{ACK} flag  \end{tabular}
& \begin{tabular}[c]{@{}l@{}} Warning \\ Warning  \end{tabular}\\ \midrule
TPKT & \begin{tabular}[c]{@{}l@{}}\lstt{Version} field must be 3 \\ \lstt{Reserved} field must be 8 bits   \end{tabular} & \begin{tabular}[c]{@{}l@{}} None \\ None  \end{tabular}\\ \midrule
PPP & \begin{tabular}[c]{@{}l@{}}\lstt{Code} field must be between 1-11\\ \lstt{Length} field must be 1 octet, at least size 4\\ \lstt{Data} must be constrained by \lstt{Length} \end{tabular} & \begin{tabular}[c]{@{}l@{}}None \\ None \\  None \end{tabular} \\ \midrule
RTP & \lstt{Version} must be 2 & None \\ \midrule
OSPF & \begin{tabular}[c]{@{}l@{}}\lstt{Version} field must not be 3 \\ \lstt{Reserved} field must be 0 \\ Header length must be 16 bytes  \end{tabular}
& \begin{tabular}[c]{@{}l@{}} None \\ None \\ None  \end{tabular}\\  \bottomrule
\end{tabular}%
}
\caption{Constraints specified in the RFC, that wireshark does not enforce. }
\label{tab:disagreement}
\end{table}

%% file: body/rqs/rq2.tex
\subsection{Distinguishing Candidates with Differential Testing}
\label{sec:rq2}

In many cases, \ourtool produces multiple specifications that are compatible with the test set. In this section, we aim to answer the following question:

\noindent\textbf{RQ2:} \rqb

We make use of $\symbTestGen$ to help us answer this question, in particular its \emph{differential testing} feature, to find tests that distinguish specifications or prove them semantically equivalent. Distinguishing tests, if any, can be surfaced to the user, along with feedback from $\symbTestGen$ localizing semantic differences in 3D specifications, to help the user identify their desired specification.

Table \ref{tab:rq2} shows protocols for which \ourtool generates multiple candidate specifications, and the results of $\symbTestGen$'s differential testing between every pair of candidates, with a description of the differences found, if any. Out of 8 protocols, 7  have at least two semantically distinct specifications, whereas for VXLAN the two candidates are semantically equivalent.

For example, for ARP, \ourtool generates 3 candidate specifications, two which are semantically equivalent, whereas the third mistakenly adds a field to the end of the ARP header \ls`UINT8 remainder[:consume-all]`. This field consumes the rest of the data in the packet and is often used for optional variable length fields. The ARP RFC 826 does not describe such a field, and the specification is incorrect but passes the test set because there is no negative-label test for which there is additional data at the end of the ARP header.

\ourtool generates two candidate specifications for IPV4, and $\symbTestGen$ find a test that distinguishes them. One specification has a field: \ls`UINT16BE flags:3 { flags == Reserved0` \ls`|| flags == DF || flags == MF }`, where \ls`Reserved0, DF, MF` are equal to \ls`0`, \ls`1`, and \ls`2`, which is consistent with the RFC. The second specification has the same field as \ls`UINT16BE Flags:3` and does not enforce constraints on the value. While these constraints should be added to be consistent with the RFC, the test set did not contain a test violating this constraint, thus both specifications were able to pass the test set. 
  
In both the cases of ARP and IPV4, $\symbTestGen$ labels the one specification as more permissive than the other, e.g. for IPV4 every test that passes on the first specification also passes on the second, but not the other way around. For both of these cases, the stricter specification correctly implements the RFC. Thus, a user of \ourtool could decide to always accept the stricter specification as a pruning heuristic between candidates.

\RS{2}{Multiple distinct specifications may be produced by \ourtool for a single protocol, and the degree to which they diverge is dependent on the quality and coverage of the test suite on which they are evaluated. \symbTestGen helps by finding differentiating tests or by grouping equivalent candidates, allowing users to focus on a semantic differences exhibited by concrete test cases.}

\begin{table}[]
\centering
\resizebox{\columnwidth}{!}{%
\begin{tabular}{@{}lccl@{}}
\toprule
\textbf{Protocol} & \begin{tabular}{c} \textbf{\# Candidates} 
\end{tabular} & \begin{tabular}{c} \textbf{\# Distinct} \\ \textbf{Candidates}
\end{tabular} & \textbf{Divergent Fields}\\ \midrule
UDP & 5 & 2 &  Optional \lstt{Data[:consume-all]} field\\ \midrule
IPV4 & 2 & 2 & Additional constraints on \lstt{Flag} values\\ \midrule
VXLAN & 2 & 1 & None \\ \midrule
DHCP & 2 & 2 &  \begin{tabular}{l} \lstt{options} field length \\ Additional constraints on \lstt{Flags} field \end{tabular}\\ \midrule
ARP & 3 & 2 & Incorrect additional \lstt{remainder[:consume-all]} field \\ \midrule
NTP & 3 & 3 & Additional constraints on LeapIndicator, Status, Type fields \\ \midrule
IGMP & 5 & 3 & Optional \lstt{OtherFields[:consume-all]}\\ \midrule
TCP & 2 & 2 & Constraints on \lstt{Options} field \\

 \bottomrule  
\end{tabular}%
} \\
\caption{Semantically distinct candidate specifications}
\label{tab:rq2}
\end{table}

%% file: body/rqs/rq3.tex
\subsection{\ourtool vs. Human Written Specs}
\label{sec:rq3}

The authors of EverParse provide specifications for 7 out of the 20 protocols we ran \ourtool on. In this section, we ask:

\noindent\textbf{RQ3:} \rqc

As before, we use $\symbTestGen$'s differential testing to semantically compare \ourtool's specifications to  the handwritten ones, with the results in Table \ref{tab:rq3}. For IPV6 and Ethernet, $\symbTestGen$ proves that the specifications are equivalent, though syntactically distinct.

For UDP, ICMP, and VXLAN, we use $\symbTestGen$ to identify tests that distinguish the handwritten and generated specifications. In all three cases, the root cause is incorrect or missing constraints in the handwritten specifications, demonstrating that even experts make mistakes when interpreting RFCs as 3D, and that \ourtool can help in ensuring consistency with RFCs. For UDP, the handwritten specification is missing a constraint on the \lstt{Length} field that exists in the \ourtool specification. For ICMP, the \lstt{Unused Bytes} field is too short, misinterpreting the 32 bytes as 32 bits. Similarly, in VXLAN, the \lstt{VXLanID} is two bytes short. After correcting the handwritten specification, $\symbTestGen$ proves them equivalent to the \ourtool generated specifications. Pull requests with the revised specifications were merged into EverParse for all three protocols.

On the other hand, for TCP and IPV4, the \ourtool specification is  missing constraints that exist in the handwritten specification. For TCP, although the produced specification passes on the set of tests, it is underconstrained and does not include a condition checking if the \lstt{SYN} flag is set to 1 and it does not implement all possible constraints for different \lstt{Max Segment Size} payloads. Instead it includes size constraints that are general to all options. 

For IPV4, there are missing constraints on the \ls`IHL`, \ls`TotalLength`, and \ls`Options` fields, which indicates that tests labeled by Wireshark do not capture these constraints. We explore whether \ourtool would be able to generate an equivalent specification, if it had access to a set of tests that can capture the missing constraints. To do this, we use $\symbTestGen$ to generate positive and negative tests from the handwritten specification, guaranteeing that tests exercising the constraints will be included in the test set. With these tests, \ourtool is able to produce two passing specifications that contain the missing constraints, after an average of 11 syntax and 4 packet refinements. The generated specifications are both semantically equivalent to the handwritten specification.

\RS{3}{\ourtool is able to produce 3D specifications semantically equivalent to human written 3D. In addition, using our framework we were able to uncover three bugs in existing handwritten 3D code for UDP, ICMP, and VXLAN, highlighting the difficulties in translating RFCs into correct implementations. }

\begin{table}[h!]
\centering
\resizebox{\columnwidth}{!}{%
\begin{tabular}{@{}l|c|l|l@{}} \toprule
Protocol & Equivalent? & Root Cause Divergence & After H.S. Fix\\ \midrule
UDP &  \tikzxmark  & H.S. Missing constraint on Length field & \checkmark \\
ICMP & \tikzxmark & H.S. UNUSED\_BYTES type too short  & \checkmark \\
VXLAN & \tikzxmark & H.S. VXLanID field too short & \checkmark \\
IPV6 & \checkmark & None & n/a \\
IPV4 & \tikzxmark & G.S Missing value constraints on IHL, TotalLength &  n/a \\
Ethernet & \checkmark & None & n/a \\
TCP & \tikzxmark & G.S. Missing constraints on options payload & n/a \\ \bottomrule
\end{tabular}%
}
\caption{Comparison of \ourtool generated specifications  to handwritten specifications (denoted H.S.). We list the cause(s) of divergence, and where applicable, we correct the handwritten specification. }
\label{tab:rq3}
\end{table}

%% file: body/related.tex
\section{Related Work}

LLMs have enabled generating code from informal natural language requirements, and have shown ability to generate human like code on benchmark problems~\cite{chen2021evaluating, austin2021program} -- however, they come with no guarantees, and have been known to contain bugs and security errors~\cite{pearce22asleep}. Alphacode~\cite{alphacode_2022} and CodeT~\cite{codet_2022} have used tests to cluster and rank generated code to improve the empirical accuracy on benchmarks; however they do not add trust to the generated code as the natural language does not impose any correctness checks.

On the other hand, classical program synthesis~\cite{manna80} formulates the problem of generating code that meets a formal specification. However, these techniques are limited due to lack of availability of formal specifications, along with the intractable theoretical complexity of the synthesis. Lately, program synthesis has been applied in restricted domains with input-output examples as specifications~\cite{gulwani_2017}, constrained by restrictions on syntax (e.g., SyGuS~\cite{sygus13}). These restrictions make it difficult to apply them for new domains with formal guarantees. Office Domain Specific Language (ODSL)~\cite{gandhi2023natural} has been proposed as an intermediate layer for LLMs to translate natural language user commands to programs over  Office APIs. Although it shares our motivation for using 3D as a DSL, generated programs from ODSL do not have any formal guarantees since the generated programs lack a formal notion of correctness and there is no symbolic encoding of programs in ODSL into a logical formula. 

Closer to our setting, Ticoder~\cite{lahiri2023interactive} uses LLMs to partially formalize user-intent as tests. Unlike \ourtool, TiCoder requires a user in the loop to validate each test, relies on an LLM-based test generation that cannot be as exhaustive as our symbolic technique and cannot provide any formal guarantees on the generated code.  Endres et al.~\cite{endres2023formalizing} generate declarative postconditions in Java and Python using LLMs and evaluate the quality of specifications offline using validation tests, but do not generate tests or verified code. Misu et al.~\cite{rakib2024towards} generate formal specifications and code that satisfies such specifications in Dafny programming language using LLMs, but there is no automation in helping the user establish the correctness of the specifications. 

All these approaches are evaluated for a simple setup where the requirements are present as a few line docstrings, and do not require problem decomposition or the translation of requirements from complex documents such as RFCs.
SAGE~\cite{Jane2021Sage} uses natural language processing (NLP) techniques to translate informal requirements in RFCs into protocol implementations semi-automatically. SAGE extracts and surfaces ambiguities in RFCs through an intermediate logical form, that are resolved by the user, before generating code. Unlike \threedGen, SAGE can generate protocol implementation in addition to the parser; however, the generated code may have functional and security bugs, as it lacks formal specifications. Extending \threedGen and the 3D language to support full protocol implementation would be interesting avenue for future work.

%% file: body/7.conclusion.tex
\section{Conclusion}
\label{sec:conclusion}

Programming in natural language using AI is a powerful new capability. But, for AI-based program synthesizers to be truly useful, they must also be trustworthy. We believe coupling AI programming assistants with symbolic tools to support intent formalization and refinement, as well guarantees about generated outputs, is a key step towards fully realizing their potential. We have explored this idea, showing it is possible to synthesize verified binary format parsers from specification documents using AI agents, while providing symbolic test-case generators to help both humans and AIs confirm and refine intent, and verification tools to ensure that intent is preserved down to executable C code. As a next step, we plan to evaluate our approach through user studies to assess whether tools like \ourtool more easily enable humans to author correct-by-construction programs in new DSLs.

%% file: body/8.Appendix.tex
\subsection{Branch coverage with 3DTestGen}

In this supplement, we briefly cover how 3DTestGen produces tests that achieve branch coverage.

Recall the $message$ parser from Section III of the paper. It has has two branches and four paths: fail to parse the first byte; fail because the first byte is less than 42; fail to parse the second byte; success.

Thus, to query Z3 for models that achieve a form of branch coverage, we instrument our encoding with branch tags---in practice, 3DTestGen does not instrument every branch, heuristically only picking branches of interest, focusing on those that involve value constraints and $casetype$s.

Next, we introduce a global uninterpreted function, $branch-trace$, representing a trace of branches to be taken. The $State$ argument to an encoded parser also contains a $branch-index$ component which records a position in the branch trace. For a given branch trace, the encoding forces the parser to follow that trace of branches. By querying Z3 while fixing the branch trace, we obtain models of the $Input$ that also are compatible with the trace, i.e., the $Input$ is a byte sequence that forces the parser to follow the branch trace. Then, by simply enumerating the traces up to a user-specified branch depth in a depth-first fashion, we get Z3 to generate a diversity of $Input$ models that achieve branch coverage.

We show a fragment of the $parse\_message$ parser encoding augmented with branch tags below. Note how each branch of the constraint check for $x > 42$ is taken only if the corresponding branch tag in the trace also permits it (0 if the constraint holds, 1 if it does not.)

\begin{lstlisting}[language=SMT2]
(define-fun parse-message ((s0 State)) State
  (let ((s1 (parse-uint8 s0)))
      (if (has-failed s1) s1
        (if (and (> (return-value s1) 42)
                 (= 0 (branch-trace (branch-index s1))))
            (parse-uint8 (incr-branch-index s1))
            (if (and (not (> (return-value s1) 42))
                     (= 1 (branch-trace (branch-index s1))))
\end{lstlisting}

Now, to generate models that explore a given trace of branches, we add the following assertions to the query to constrain the prefix of the branch trace recorded by the parser.

\begin{lstlisting}[language=SMT2]
(assert (= (branch-index init) 0)) ;; start from index 0.
;; take the first branch in message.
(assert (and (= (branch-trace 0) 0))) 
;; make at least as many choices as branch-depth.
(assert (>= (branch-index (parse-message init)) 
              branch-depth)) 
\end{lstlisting}

\begin{figure*}
\centering
\includegraphics[width=0.85\textwidth]{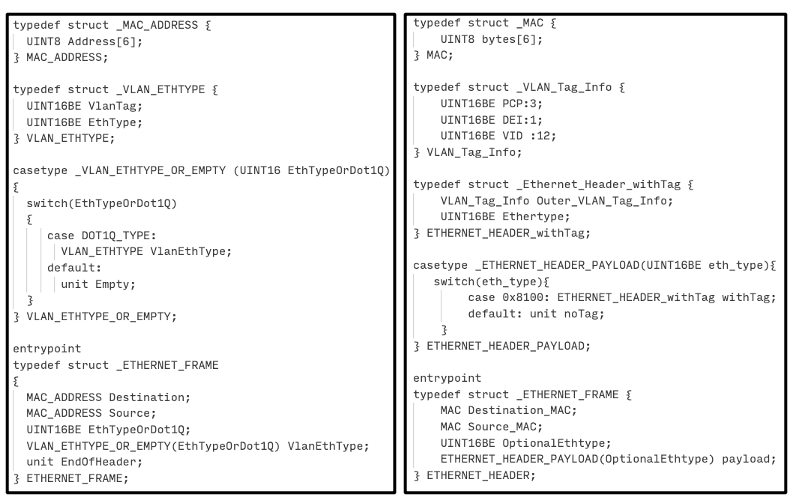}
    \caption{Semantically equivalent 3D specification for Ethernet II frames in VXLAN. (Left) Handwritten Specification, (Right) \ourtool Specification.}
    \label{fig:eth}
\end{figure*}

\begin{figure*}
\centering
\includegraphics[width=0.85\textwidth]{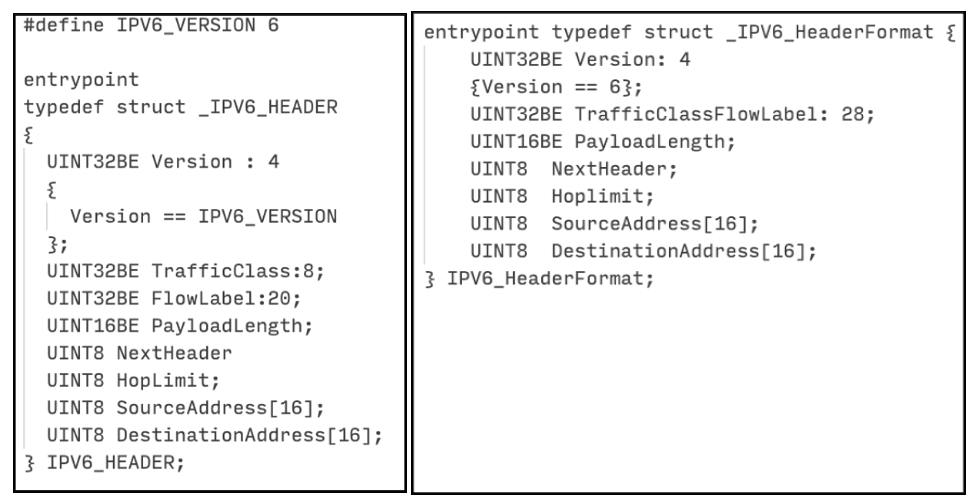}
    \caption{Semantically equivalent 3D specification for IPV6. (Left) Handwritten Specification, (Right) \ourtool Specification.}
    \label{fig:ipv6}
\end{figure*}

\subsection{Comparing code produced by 3DGen to Handwritten Specifications}

We report the high level characteristics of code produced by \ourtool in Table~\ref{tab:loc}. For each characteristic (Lines of code (LOC), number of constraints, etc.) we report the mean across all valid specifications generated by \ourtool in the first column. In the last two columns, we compare characteristics of the generated specification to the correct handwritten specification, if applicable, across all protocols. Table~\ref{tab:breakdown} contains the full breakdown of metrics for each protocol. 

In general, we observe that handwritten specifications have on average, more lines of code (ignoring comments and empty lines), than code generated by \ourtool. This could be due to formatting and stylistic choices. For example, in Figure~\ref{fig:ipv6}, the handwritten specification adds extra lines around the version constraint, and extracts the version number into a defined variable, where the \ourtool spec uses a constant.

While the mean is the same, we observe that the number of bitfields can vary between the generated and handwritten specifications. An example of this can be seen in Figure~\ref{fig:ipv6}, where the handwritten specification divides the $TrafficClass$ and $FlowLabel$ fields. The \ourtool specification combined both of these fields into one $TrafficClassFlowLabel$ field of the appropriate size. There is no constraint on the value of this field in either specification, therefore, both specifications are semantically equivalent. On the other hand, Figure~\ref{fig:eth} shows an example of semantically equivalent specification where \ourtool represented the $VlanTag$ field with three bitfield variables $PCP$, $DEI$, and $VID$. 

Interestingly, both \ourtool and the handwritten specifications contain the same number of casetypes, with a mean of 0.42 across the seven specifications for which a handwritten spec exists. The number of constraints in the \ourtool specification is lower than that of the handwritten specifications, which follows from observations in RQ3, Table~\ref{tab:rq3}, where the IPV6 and TCP specifications are missing constraints.

\begin{table}[]
\centering
\resizebox{=0.8\columnwidth}{!}{%
\begin{tabular}{@{}l|l|ll@{}}
\toprule
Metric (mean) & All Protocols & \ourtool & H.S. \\ \midrule
LOC (3D) & 26.2 & 33.14 & 60 \\
Fields & 15.7 & 22.5 & 24.1 \\
Constraints & 3.4 & 4 & 5.14 \\
Casetypes & 0.3 & 0.42 & 0.42 \\
Structs & 2.75 & 4.14 & 4.14 \\
Bitfields & 3.35 & 4.71 & 4.71 \\
Enum & 0.15 & 0.00 & 0.42 \\
Consume-all & 0.71 & 0.42 & 0.28 \\  \bottomrule
\end{tabular}%
}
\caption{Highlevel characteristics of code produced by 3DGen. Mean is reported across all specifications. For the subset for which a handwritten spec exists, the last two columns compare characteristics between generated specifications (G.S.) and handwritten specifications (H.S.). }
\label{tab:loc}
\end{table}

\subsection{Syntactic Characteristics of code produced by \ourtool}

\begin{table*}[h!]
\centering
\resizebox{=0.9\linewidth}{!}{%
\begin{tabular}{@{}rl|ccccccccc@{}}
\toprule
\textbf{\#} & \textbf{Protocol} &  \textbf{LOC (C)}& \textbf{LOC (3D)}  & \textbf{Fields}  &  \textbf{Constraints}  & \textbf{Casetypes} &\textbf{Structs}  & \textbf{Bitfields} & \textbf{Enum} & \textbf{Consume-all}   \\ \midrule
1 & UDP & 158 & 7 & 4 & 1 & 0& 1 & 0 & 0&0      \\
 & H.S. & 158 & 8 & 4 & 1 & 0 & 1 & 0 &0 &0  \\ \midrule
2 & ICMPv4 & 1208 & 99 & 72 & 11 & 1& 12& 0 & 0& 2   \\
 & H.S. & 2119 & 164 & 61 & 6 & 1 & 13 & 0 & 3 & 1 \\ \midrule
3 & VXLAN & 296 & 16 & 10 & 7 & 0&3 & 5 &0 & 0  \\
 & H.S. & 266 & 10 & 7 & 6 & 0 & 1& 3 &0 &0  \\ \midrule
4 & IPV6 & 341 & 10 &7 &1 & 0& 1& 2 &0 &0   \\
 & H.S. & 427 & 16 & 7 & 1 & 0 & 1& 3 & 0& 0 \\ \midrule
5 & IPV4 & 544 & 30& 25 &5 & 0& 3& 15& 0& 0    \\
 & H.S. & 534 & 35& 15 & 5 & 0 & 2& 6 & 0& 0 \\ \midrule
 
6 & TCP & 615 & 45& 27 & 3 & 1& 5& 8 & 0& 1    \\
 & H.S. & 1445 & 161& 64 & 17 & 1 & 8& 21 & 0& 1 \\ \midrule
7 & Ethernet & 352 & 25 &13 & 0 & 1& 4& 3 & 0& 0  \\ 
 & H.S. & 353 & 26 & 11 & 0 & 1 & 3 &0  & 0& 0 \\ \midrule \midrule
8 & GRE & 309 & 22 & 12 & 0 & 1 & 3 & 3 & 0 &1     \\
9 & IGMPv2 & 223 & 10& 4 & 4 & 0&1 & 0 & 0& 0    \\
10 & DHCP & 838 & 25 & 18 & 5 & 0 & 3 & 0 & 0 & 0  \\
11 & DCCP & 491 & 38 & 17& 0 & 1& 3& 4& 1&  1  \\
12 & ARP & 530 & 12& 10 & 0 & 0&1 & 0 &0 &0   \\
13 & NTP & 497 & 35& 15 & 3 & 0& 2& 3& 0& 1   \\
14 &  NBNS & 416 & 17& 15& 14 & 0&1 & 10 & 0& 0  \\
15 & NSH & 486 & 24 & 16 & 1 & 0& 4 & 8 & 0& 1    \\ 
16 & TFTP & 304 & 43 &15 & 7 & 1& 1& 0& 0& 4    \\
17 & RTP & 320 & 17& 13& 1 &0 & 2&6 &0 & 1   \\
18 & PPP & 270 & 20 & 6 & 1 & 0&1 & 0 & 1& 1   \\ 
19 & TPKT & 232 & 6 & 4& 3 & 0& 1& 0& 0&  0 \\
20 & OSPF & 351 &  23 & 11 & 1 & 0&3 & 0& 1& 1  \\ 

\bottomrule \\
\end{tabular}
}
\caption{Syntactic metrics of 3D code produced by \ourtool and the handwritten specification (H.S.) for each of the 20 protocols used in this work.}
\label{tab:breakdown}
\end{table*}

Results of RQ3 in this work, as well as results from related work in the domain, confirm that accurately translating nuanced and often ambiguous RFC specifications into correct code by hand is an error prone task and often laborious task.
Table~\ref{tab:breakdown} contains several syntactic metrics of the code generated by \ourtool for each protocol. Though these syntactic metrics may not serve as an ideal proxy for the complexity of a protocol's data format specification, they do give some insight into the relative complexity and structural elements of the generated code. 

We report both the lines of 3D code generated by \ourtool as well as the lines of C code automatically generated from the 3D specification by Everparse. Overall, we observe a modest range of lines of 3D code generated for the 20 selected protocols, \ourtool generated as little as 6 (TPKT) and at most 99 (ICMP) lines of code. Intuitively, the number of fields follows overall lines of code, ranging from 4 (IGMP, TPKT) to 72 (ICMP).  Although we observe only a modest numbers of lines of 3D code generated, the lines of auto-generated C code to correctly parse these specifications is considerably larger---we comment more on this shortly.

The number of structs, bitfields, enums, and consume-all types used varies based on the protocol specification. The number of value constraints can indicate increasing nuance in the data specification. For example, NBNS spans 17 lines of code but contains 14 constraints, as measured by boolean operators. This is because of the 10 bitfields, 8 have value constraints defined in the NBNS RFC 1002. The \texttt{NM\_FLAGS} field defines value constraints for each flag of either 0 or 1. The casetypes field appears to be a reliable indicator if the protocol contains multiple message types, however for the protocols studied there is never more than one casetype needed to specify the data type. 

Interestingly, the number of lines of C code produced for each of the formats is several factors larger than the size of the 3D specification, for several reasons. First, designed as as declarative specification language, 3D is inherently more compact than imperative C code that implements a parser, e.g., the parser implementation has to repeatedly check if there is enough space left in the input buffer to parse the next field. Further, the C code includes various features that are important for a practical parser, including error handling logic---such error handling is not present in individual 3D specifications and is instead baked into the definition of the 3D language. This points to the benefit of using a compact specification language as a target for AI-generation---many features can be incorporated into the language definition, rather than having the programmer specify them repeatedly for every program. Directly generating, say, 1K lines of C code for an ICMP parser, even if it could be AI-generated, would pose a difficult program verification problem; using a DSL like 3D enables analysis such as symbolic test-case generation to systematically test and refine AI generated code, and correct-by-construction code generation ultimately yields verified C code.

%% file: main.bbl
\begin{thebibliography}{10}
\providecommand{\url}[1]{#1}
\csname url@samestyle\endcsname
\providecommand{\newblock}{\relax}
\providecommand{\bibinfo}[2]{#2}
\providecommand{\BIBentrySTDinterwordspacing}{\spaceskip=0pt\relax}
\providecommand{\BIBentryALTinterwordstretchfactor}{4}
\providecommand{\BIBentryALTinterwordspacing}{\spaceskip=\fontdimen2\font plus
\BIBentryALTinterwordstretchfactor\fontdimen3\font minus \fontdimen4\font\relax}
\providecommand{\BIBforeignlanguage}[2]{{%
\expandafter\ifx\csname l@#1\endcsname\relax
\typeout{** WARNING: IEEEtran.bst: No hyphenation pattern has been}%
\typeout{** loaded for the language `#1'. Using the pattern for}%
\typeout{** the default language instead.}%
\else
\language=\csname l@#1\endcsname
\fi
#2}}
\providecommand{\BIBdecl}{\relax}
\BIBdecl

\bibitem{nail14osdi}
\BIBentryALTinterwordspacing
J.~Bangert and N.~Zeldovich, ``Nail: A practical tool for parsing and generating data formats,'' in \emph{11th USENIX Symposium on Operating Systems Design and Implementation (OSDI 14)}.\hskip 1em plus 0.5em minus 0.4em\relax Broomfield, CO: USENIX Association, Oct. 2014, pp. 615--628. [Online]. Available: \url{https://www.usenix.org/conference/osdi14/technical-sessions/presentation/bangert}
\BIBentrySTDinterwordspacing

\bibitem{everparse19usenix}
T.~Ramananandro, A.~Delignat-Lavaud, C.~Fournet, N.~Swamy, T.~Chajed, N.~Kobeissi, and J.~Protzenko, ``Everparse: Verified secure zero-copy parsers for authenticated message formats,'' in \emph{Proceedings of the 28th USENIX Conference on Security Symposium}, ser. SEC'19.\hskip 1em plus 0.5em minus 0.4em\relax USA: USENIX Association, 2019, p. 1465–1482.

\bibitem{everparse3d22pldi}
\BIBentryALTinterwordspacing
N.~Swamy, T.~Ramananandro, A.~Rastogi, I.~Spiridonova, H.~Ni, D.~Malloy, J.~Vazquez, M.~Tang, O.~Cardona, and A.~Gupta, ``Hardening attack surfaces with formally proven binary format parsers,'' in \emph{Proceedings of the 43rd ACM SIGPLAN International Conference on Programming Language Design and Implementation (PLDI '22), June 13--17, 2022, San Diego, CA, USA}, 2022. [Online]. Available: \url{https://www.fstar-lang.org/papers/EverParse3D.pdf}
\BIBentrySTDinterwordspacing

\bibitem{openai2023gpt4}
OpenAI, ``Gpt-4 technical report,'' 2023.

\bibitem{lowstar}
\BIBentryALTinterwordspacing
J.~Protzenko, J.-K. Zinzindohou\'e, A.~Rastogi, T.~Ramananandro, P.~Wang, S.~{Zanella-B\'eguelin}, A.~Delignat-Lavaud, C.~Hritcu, K.~Bhargavan, C.~Fournet, and N.~Swamy, ``Verified low-level programming embedded in {F*},'' \emph{{PACMPL}}, vol.~1, no. {ICFP}, pp. 17:1--17:29, Sep. 2017. [Online]. Available: \url{http://arxiv.org/abs/1703.00053}
\BIBentrySTDinterwordspacing

\bibitem{yao2022react}
S.~Yao, J.~Zhao, D.~Yu, N.~Du, I.~Shafran, K.~Narasimhan, and Y.~Cao, ``React: Synergizing reasoning and acting in language models,'' \emph{arXiv preprint arXiv:2210.03629}, 2022.

\bibitem{gao2023retrieval}
Y.~Gao, Y.~Xiong, X.~Gao, K.~Jia, J.~Pan, Y.~Bi, Y.~Dai, J.~Sun, and H.~Wang, ``Retrieval-augmented generation for large language models: A survey,'' \emph{arXiv preprint arXiv:2312.10997}, 2023.

\bibitem{xi2023rise}
Z.~Xi, W.~Chen, X.~Guo, W.~He, Y.~Ding, B.~Hong, M.~Zhang, J.~Wang, S.~Jin, E.~Zhou \emph{et~al.}, ``The rise and potential of large language model based agents: A survey,'' \emph{arXiv preprint arXiv:2309.07864}, 2023.

\bibitem{wang2023survey}
L.~Wang, C.~Ma, X.~Feng, Z.~Zhang, H.~Yang, J.~Zhang, Z.~Chen, J.~Tang, X.~Chen, Y.~Lin \emph{et~al.}, ``A survey on large language model based autonomous agents,'' \emph{arXiv preprint arXiv:2308.11432}, 2023.

\bibitem{du2023improving}
Y.~Du, S.~Li, A.~Torralba, J.~B. Tenenbaum, and I.~Mordatch, ``Improving factuality and reasoning in language models through multiagent debate,'' \emph{arXiv preprint arXiv:2305.14325}, 2023.

\bibitem{qian2023communicative}
C.~Qian, X.~Cong, C.~Yang, W.~Chen, Y.~Su, J.~Xu, Z.~Liu, and M.~Sun, ``Communicative agents for software development,'' \emph{arXiv preprint arXiv:2307.07924}, 2023.

\bibitem{wu2023autogen}
Q.~Wu, G.~Bansal, J.~Zhang, Y.~Wu, S.~Zhang, E.~Zhu, B.~Li, L.~Jiang, X.~Zhang, and C.~Wang, ``Autogen: Enabling next-gen llm applications via multi-agent conversation framework,'' \emph{arXiv preprint arXiv:2308.08155}, 2023.

\bibitem{SMTLIB}
C.~Barrett, P.~Fontaine, and C.~Tinelli, ``{The Satisfiability Modulo Theories Library (SMT-LIB)},'' \url{https://smtlib.cs.uiowa.edu/}, 2016.

\bibitem{de2008z3}
L.~De~Moura and N.~Bj{\o}rner, ``Z3: An efficient smt solver,'' in \emph{Tools and Algorithms for the Construction and Analysis of Systems: 14th International Conference, TACAS 2008, Held as Part of the Joint European Conferences on Theory and Practice of Software, ETAPS 2008, Budapest, Hungary, March 29-April 6, 2008. Proceedings 14}.\hskip 1em plus 0.5em minus 0.4em\relax Springer, 2008, pp. 337--340.

\bibitem{chen2021evaluating}
M.~Chen, J.~Tworek, H.~Jun, Q.~Yuan, H.~P. d.~O. Pinto, J.~Kaplan, H.~Edwards, Y.~Burda, N.~Joseph, G.~Brockman \emph{et~al.}, ``Evaluating large language models trained on code,'' \emph{arXiv preprint arXiv:2107.03374}, 2021.

\bibitem{austin2021program}
J.~Austin, A.~Odena, M.~Nye, M.~Bosma, H.~Michalewski, D.~Dohan, E.~Jiang, C.~Cai, M.~Terry, Q.~Le \emph{et~al.}, ``Program synthesis with large language models,'' \emph{arXiv preprint arXiv:2108.07732}, 2021.

\bibitem{pearce22asleep}
H.~Pearce, B.~Ahmad, B.~Tan, B.~Dolan-Gavitt, and R.~Karri, ``Asleep at the keyboard? assessing the security of github copilot’s code contributions,'' in \emph{2022 IEEE Symposium on Security and Privacy (SP)}, 2022, pp. 754--768.

\bibitem{alphacode_2022}
\BIBentryALTinterwordspacing
Y.~Li, D.~Choi, J.~Chung, N.~Kushman, J.~Schrittwieser, R.~Leblond, T.~Eccles, J.~Keeling, F.~Gimeno, A.~D. Lago, T.~Hubert, P.~Choy, C.~d.~M. d'Autume, I.~Babuschkin, X.~Chen, P.-S. Huang, J.~Welbl, S.~Gowal, A.~Cherepanov, J.~Molloy, D.~J. Mankowitz, E.~S. Robson, P.~Kohli, N.~de~Freitas, K.~Kavukcuoglu, and O.~Vinyals, ``Competition-level code generation with alphacode,'' 2022. [Online]. Available: \url{https://arxiv.org/abs/2203.07814}
\BIBentrySTDinterwordspacing

\bibitem{codet_2022}
\BIBentryALTinterwordspacing
B.~Chen, F.~Zhang, A.~Nguyen, D.~Zan, Z.~Lin, J.-G. Lou, and W.~Chen, ``Codet: Code generation with generated tests,'' 2022. [Online]. Available: \url{https://arxiv.org/abs/2207.10397}
\BIBentrySTDinterwordspacing

\bibitem{manna80}
\BIBentryALTinterwordspacing
Z.~Manna and R.~Waldinger, ``A deductive approach to program synthesis,'' \emph{ACM Trans. Program. Lang. Syst.}, vol.~2, no.~1, p. 90–121, jan 1980. [Online]. Available: \url{https://doi.org/10.1145/357084.357090}
\BIBentrySTDinterwordspacing

\bibitem{gulwani_2017}
\BIBentryALTinterwordspacing
S.~Gulwani, O.~Polozov, and R.~Singh, ``Program synthesis,'' \emph{Found. Trends Program. Lang.}, vol.~4, no. 1-2, pp. 1--119, 2017. [Online]. Available: \url{https://doi.org/10.1561/2500000010}
\BIBentrySTDinterwordspacing

\bibitem{sygus13}
\BIBentryALTinterwordspacing
R.~Alur, R.~Bod{\'{\i}}k, G.~Juniwal, M.~M.~K. Martin, M.~Raghothaman, S.~A. Seshia, R.~Singh, A.~Solar{-}Lezama, E.~Torlak, and A.~Udupa, ``Syntax-guided synthesis,'' in \emph{Formal Methods in Computer-Aided Design, {FMCAD} 2013, Portland, OR, USA, October 20-23, 2013}.\hskip 1em plus 0.5em minus 0.4em\relax {IEEE}, 2013, pp. 1--8. [Online]. Available: \url{https://ieeexplore.ieee.org/document/6679385/}
\BIBentrySTDinterwordspacing

\bibitem{gandhi2023natural}
A.~Gandhi, T.~Q. Nguyen, H.~Jiao, R.~Steen, and A.~Bhatawdekar, ``Natural language commanding via program synthesis,'' 2023.

\bibitem{lahiri2023interactive}
\BIBentryALTinterwordspacing
S.~K. Lahiri, S.~Fakhoury, A.~Naik, G.~Sakkas, S.~Chakraborty, M.~Musuvathi, P.~Choudhury, C.~von Veh, J.~P. Inala, C.~Wang, and J.~Gao, ``Interactive code generation via test-driven user-intent formalization,'' \emph{CoRR}, vol. abs/2208.05950, 2023. [Online]. Available: \url{https://doi.org/10.48550/arXiv.2208.05950}
\BIBentrySTDinterwordspacing

\bibitem{endres2023formalizing}
\BIBentryALTinterwordspacing
M.~Endres, S.~Fakhoury, S.~Chakraborty, and S.~K. Lahiri, ``Formalizing natural language intent into program specifications via large language models,'' \emph{CoRR}, vol. abs/2310.01831, 2023. [Online]. Available: \url{https://doi.org/10.48550/arXiv.2310.01831}
\BIBentrySTDinterwordspacing

\bibitem{rakib2024towards}
M.~Md~Rakib Hossain~Misu, C.~V. Lopes, I.~Ma, and J.~Noble, ``Towards ai-assisted synthesis of verified dafny methods,'' 2024.

\bibitem{Jane2021Sage}
\BIBentryALTinterwordspacing
J.~Yen, T.~L\'{e}vai, Q.~Ye, X.~Ren, R.~Govindan, and B.~Raghavan, ``Semi-automated protocol disambiguation and code generation,'' in \emph{Proceedings of the 2021 ACM SIGCOMM 2021 Conference}, ser. SIGCOMM '21.\hskip 1em plus 0.5em minus 0.4em\relax New York, NY, USA: Association for Computing Machinery, 2021, p. 272–286. [Online]. Available: \url{https://doi.org/10.1145/3452296.3472910}
\BIBentrySTDinterwordspacing

\end{thebibliography}
